\documentclass[structabstract]{aa}
  
\usepackage{graphicx}
\usepackage{float}
\usepackage{subfig}
\usepackage{txfonts}
\usepackage{calc}

\usepackage{color}

\begin{document}
   \title{Eclipsing Spotted Giant Star With K2 and Historical Photometry}

   \author{K. Ol\'ah\inst{1}, S. Rappaport\inst{2}, T. Borkovits\inst{3,1}, T. Jacobs\inst{4}, D. Latham\inst{5}, A. Bieryla\inst{5}, I.B. B\'{i}r\'o\inst{3}, J. Bartus\inst{1,6}, Zs. K\H ov\'ari\inst{1}, K. Vida\inst{1},  A. Vanderburg\inst{7}, D. LaCourse\inst{8}, I. Cs\'anyi\inst{3}, G. \'A. Bakos\inst{9,10,11}, W. Bhatti\inst{9}, Z. Csubry\inst{9}, J. Hartman\inst{9}, M. Omohundro\inst{12} 
                          \fnmsep\thanks{Data presented in this paper are based on observations obtained at the
HAT station at the Submillimeter Array of SAO, and the HAT station at
the Fred Lawrence Whipple Observatory of SAO.}
      }  

 \institute{Konkoly Observatory, Research Centre for Astronomy and
Earth Sciences, Hungarian Academy of Sciences, H-1121 Budapest, Konkoly Thege M. u. 15-17, Hungary\\
              \email{olah@konkoly.hu}
\and
Department of Physics, and Kavli Institute for Astrophysics and Space Research, Massachusetts Institute of Technology, Cambridge, MA 02139, USA
\and
Baja Astronomical Observatory of Szeged University, H-6500 Baja, Szegedi \'ut, Kt. 766, Hungary
\and
Amateur Astronomer, 12812 SE 69th Place Bellevue, WA 98006, USA
\and
Harvard-Smithsonian Center for Astrophysics, Cambridge, MA 02138, USA
\and
Leibniz-Institute for Astrophysics Potsdam (AIP), An der Sternwarte 16, 14482, Potsdam, Germany
\and
Department of Astronomy, The University of Texas at Austin, 2515 Speedway, Stop C1400, Austin, TX 78712 
NASA Sagan Fellow
\and
Amateur Astronomer, 7507 52nd Place NE Marysville, WA 98270, USA
\and
Department of Astrophysical Sciences, Princeton University
\and
MTA Distinguished Guest Fellow, Konkoly Observatory
\and Packard Fellow 
\and
Citizen Scientist, c/o Zooniverse, Department of Physics, University of Oxford, Denys Wilkinson Building, Keble Road, Oxford, OX1 3RH, UK }
               
   \date{Received ..., 2018; accepted ...}

    \abstract 
{Stars can maintain their observable magnetic activity from the pre-main sequence to the tip of the red giant branch. However, the number of known active giants is much lower than active stars on the main sequence since on the giant branch the stars spend only about 10\% of their main sequence lifetime. Due to their rapid evolution it is difficult to estimate the stellar parameters of giant stars. A possibility for obtaining more reliable stellar parameters of an active giant arises when it is a member of an eclipsing binary system.}
   {We have discovered EPIC 211759736, an active spotted giant star in an eclipsing binary system during the $Kepler K2$ Campaign 5. The eclipsing nature allows us to much better constrain the stellar parameters than in most cases of active giant stars.}
   {We have combined the $K2$ data with archival HATNet and DASCH photometry, new spectroscopic radial velocity measurements, and a set of follow-up ground-based $BVR_CI_C$ photometric observations, to find the binary system parameters as well as robust spot models for the giant at two different epochs.}
   {We determined the physical parameters of both stellar components and provide a description of the rotational and long-term activity of the primary component. The temperatures and luminosities of both components were examined in the context of the Hertzsprung--Russell diagram. We find that both the primary and the secondary components deviate from the evolutionary tracks corresponding to their masses in the sense that the stars appear in the diagram at lower masses than their true masses.}
  {We further evaluate the proposition that active giants have masses that are found to be generally higher by traditional methods than are indicated by stellar evolution tracks in the HR diagram. A possible reason for this discrepancy could be a strong magnetic field, since we see greater differences in more active stars.}

   \authorrunning{Ol\'ah et al.}
   \maketitle
%

\section{Introduction}\label{intro}

Stars can maintain their observable magnetic activity from the pre-main sequence (thereafter MS) until the tip of the red giant branch. We define an `active star' as one for which we can infer from observations surface spots due to variable magnetic fields. 
Observable activity develops in those stars where there are extra circumstances to strengthen the magnetic field, and among these is rapid rotation which is maintained in binary stars via tidal synchronization. 

Strong magnetic fields have been directly observed from Stokes vectors already in many active stars (Reiners,~\cite{reiners}). In the case of pre-main sequence stars, it is known that the strong magnetic fields can suppress the convection in the spotted regions and alter the entire stellar atmosphere which, in turn, makes the derived physical parameters of the stars uncertain (Bouvier \& Bertout, \cite{bouvier}). Having similar rotation rates, red giant stars with deep convection zones can also possess surface magnetic fields maintained by the magnetic dynamo. Therefore, their atmospheric structure can also be altered by the strong magnetic fields making their astrophysical parameters more uncertain  (see Ol\'ah et al.~\cite{overactive}). In the case of large spotted areas, the observed temperature is lower and the resulting masses and ages inferred from their location on evolution tracks can be quite inaccurate.  This is why observing an active giant star in an eclipsing binary is of great value, since this makes it possible to determine the stellar masses independently. By now, direct evidence -- i.e., an interferometric map -- shows that the atmospheres of active giants can be covered with dark and bright features of possibly magnetic origin (for further details see Roettenbacher et al.~\cite{rachael1}).

The number of known active giants is much lower than active stars on the main sequence, since on the giant branch the stars evolve rapidly, spending only about 10\% of their MS lifetime there. This is supported by the results of van Doorsselaere et al. (\cite{vandoor}) who studied a sample of {\em Kepler} stars searching for flares, where the sample had 10\% giants among the group of F+G+K+M+giant stars. Of the giant stars, 3.18\% show flares---a clear sign of magnetic activity. The occurrence rate for flaring among giants is similar to their progenitor F and G stars on the MS (2.37\% and 2.96\%, respectively), whereas K+M stars have twice as high an occurrence rate for flaring of 5.87\%.

Due to their rapid evolution it is difficult to estimate the stellar parameters of giant stars. A possibility for obtaining more reliable stellar parameters of an active giant arises when it is a member of an eclipsing binary system. Although quite a number of active stars are found in eclipsing binaries, most of those are main-sequence stars or subgiants. Only about a dozen well-studied systems have a giant star as an active component, but at the moment the only well-studied {\em active} giant star that is the primary of an eclipsing binary is BE~Psc.  BE~Psc was studied in detail by Strassmeier et al.~(\cite{klausetal}) using photometry spanning 19 years, and 61 high-resolution spectra. However, it is a lucky case, since the two components have fairly similar masses (1.56 and 1.31\,~$M_{\odot}$); one star is a giant already and the other is just leaving the MS (temperatures are 4500\,K and 6300\,K). Therefore, one eclipse is deep, more than 0.15~mag., but the other eclipse was more difficult to find. Even the deep eclipse was recognized later than the discovery of the light variations of the star.

In the case of solar size, subsolar, or red-dwarf secondaries the eclipses (if they exist at all due to the inclination of the orbit) are increasingly more shallow.  A shallow eclipse in a long-period light curve (as those of the active giants with periods of tens of days) can last from several hours to days, and can be easily missed in ground-based data due to photometric uncertainties.  Another, very important factor is that these stars are usually observed  once a night with ground-based telescopes due to their long rotation periods. Even those surveys that observe a stellar field regularly and have many data points can fail to detect shallow eclipses. 

The long-cadence {\em Kepler} data have a photometric precision per sample below millimagnitudes, and thus shallow eclipses of red giants in binary systems can be more readily recognized. Excellent examples are given in Gaulme et al. (\cite{gaulmeetal2}) for 16 eclipsing systems with red giant (or subgiant) stars using {\em Kepler} photometry, together with high-resolution spectroscopy, aimed at testing how the asteroseismic scaling relations agree with those derived from dynamical modeling. The observed eclipses make it possible to determine the stellar parameters. The depths of the observed eclipses are typically a few hundredths (between $0.01-0.05$) of a magnitude, and only in 2 cases out of 16 binaries does the eclipse depth exceed 0.1 magnitude. In addition, half of the studied systems have longer periods than 100 days.  Apart from their genuine rareness arising from the rapid evolution of these stars, all this evidence suggests that the main reason for not observing more eclipsing active giants is the quality and quantity of the ground-based observations.  

Stellar activity, which can be seen in the data themselves and in the residuals from the model fits, is detected in only 8 of the 16 systems (see Fig. 3. of Gaulme et al., \cite{gaulmeetal2}). Combining the results of Table 1 from Gaulme et al. (\cite{gaulmeetal1}), and Fig.~3 and Table 4 of Gaulme et al. (\cite{gaulmeetal2}), we find 8 systems with a total out-of-eclipse flux variability of over 1\%. The higher-amplitude rotationally-modulated light curves in 4 primaries among the active giants, having over 15\% peak-to-peak flux variability, belong to the fastest rotators ($P_{\rm rot}$ $\lesssim 41$ days) of the sample; the rotational modulation periods are given in Gaulme et al. (\cite{gaulmeetal1}). For these four primaries no solar-like p-mode oscillations were detected, probably because of suppression by stellar activity (for more discussion see Gaulme et al., \cite{gaulmeetal1}).

Three very active giants in {\em non-eclipsing} binary systems (IL~Hya, XX~Tri and DM~UMa) were studied by Ol\'ah et al. (\cite{overactive}) using decades-long photometry, all of which show long-term variability with an amplitude of about a magnitude. It was found that the derived luminosities were so low that it was impossible to get reliable ages for the systems from evolutionary tracks.  This difficulty was further exacerbated  due to the high magnetic activity which prevented Ol\'ah et al. (\cite{overactive}) from getting accurate stellar parameters. One may speculate that in the case of XX~Tri, the very high amplitude rotational modulation indicates a relatively high inclination angle, and additionally, since the secondary star is not seen in the spectra, it is very likely to be a dwarf star. Therefore, even if the geometry were favorable, the eclipses would not have been observed due to the very large luminosity difference in the two stars, leaving us with an uncertain orbital inclination.

This paper reports an active red giant star, EPIC~211759736, in a doubly eclipsing binary found in Campaign 5 (GO\,5069, 2015) and re-observed in Campaign 18 (GO\,19033, 2018) of the {\em Kepler} K2 program  (Huber et al. \cite{huberetal}). The star appears first in Schmidt et al. (\cite{schmidtetal}) as a Cepheid variable, based on ASAS measurements, with a period of 36.31 days, and in the same year in Hoffman et al.~(\cite{hoffmanetal}) as a long-period variable (Cepheid or Mira) based on NSVS data, with a period of 34.483~days. These variable-star classifications were, however, incorrect. The large-amplitude long-period modulations turn out to be due to starspots on a slowly rotating star.

The eclipsing nature of the active red giant EPIC~211759736 should allow us to much better constrain the stellar parameters than in most cases of active giant stars. The binary solution is supported by new spectroscopic data. Follow-up ground-based $BVR_CI_C$ photometry was also obtained covering one stellar rotation. We make use of archival observations by ASAS (Pojmanski \cite{ASAS}) and by HATNet (Bakos et al., \cite{gazsi1}, Bakos ~\cite{gazsi3}), and also the DASCH database (Grindlay et al., \cite{grindlayetal}) which has some $\sim$900 photographic measurements of brightness spanning about 100 years.  We note that in the Gaulme et al.~(\cite{gaulmeetal2}) sample of 16 eclipsing giants in the {\em Kepler} field, only one star has supplemental ASAS data and none was found in the DASCH historical database.

EPIC\,211759736\,=\,2MASS\,08151296+1644414 has J, H, and K magnitudes of 9.516$\pm$0.022, 8.921$\pm$0.016 and 8.766$\pm$0.018 mag., respectively (Skrutskie et al.~\cite{2MASS}), and WISE W1, W2, W3, and W4 magnitudes of 8.686$\pm$0.022, 8.743$\pm$0.019, 8.631$\pm$0.025 and $\gtrsim$8.197, respectively (the AllWise catalog; Wright et al.~\cite{WISE}). 

{\it Gaia} DR2 puts the star at a distance of $1368^{+77}_{-69}$\,pc, and gives the following parameters: radial velocity is $40.6\pm4.8$\,km/sec (note that {\it Gaia} DR2 treats single-lined binaries as one star), $T_{\rm eff} = 4747\pm200$\,K, $R = 11.0\pm1\,R_\odot$, $L = 55\pm4\,L_\odot$.

The paper is structured as follows: Section 2 describes the observational data; Section 3 presents the results from photometry, Section 4 deals with the binary solution and the results of spot modeling; Section 5 discusses the results; and Section 6 gives a summary, and conclusion.

\section{Applied data}\label{obs}

\subsection{Archival Photometric data}

EPIC~211759736 has a $\sim$100 years long photometric record in the DASCH database (Grindlay et al. \cite{grindlayetal}) from scanned photographic plates taken between 1888-1989, thereby allowing us to check the cyclic nature of the active giant. 

From the public All Sky Automated Survey (ASAS) database (Pojmanski \cite{ASAS}) a 7-yr long dataset measured in the $V$-band, was downloaded and analyzed. The ASAS project automatically monitors the entire sky with wide-field instruments targeting all stars brighter than 14 magnitude, searching for, and following photometric variability.  

Data for the star EPIC~211759736 was obtained with the Hungarian-made Automated Telescope Network (HATNet; Bakos et al., \cite{gazsi1}, Bakos ~\cite{gazsi3}), using the HAT-5 and HAT-10 telescopes at the Fred Lawrence Whipple Observatory, Arizona, and the HAT-8 telescope at the Smithsonian Astrophysical Observatory's Mauna Kea site at Hawaii. Altogether, over 15,000 data points were obtained between Nov 21, 2008 and January 17, 2012 in the Sloan $r$ filter.  Data were reduced to light curves as described in, e.g., Bakos et al.~(\cite{gazsi2}).  We used the ``magnitude fitted'' values, which correct for any smooth flux changes across the field and as a function of time (e.g.~due to extinction or focus changes), with respect to a selected reference frame.  We did {\em not} use the de-correlated or trend filtered magnitude values in the light curves, as such methods can potentially distort the light curve shape for variable stars (unless done in a signal reconstruction mode, which was not available). 

\subsection{New $BVR_CI_C$ observations}

$BVR_CI_C$ observations of EPIC\,211759736 were made with the $0.5$~m telescope of Baja Observatory of the University of Szeged, located at Baja, Hungary, and equipped with an SBIG ST-6303 CCD detector. The target was observed on 20 nights between March~13 and May~19, 2018, covering almost two orbital periods; however, gaps induced by weather conditions effectively limited the coverage to one orbital period only. The usual data reduction and photometric analysis were performed using IRAF\footnote{IRAF is distributed by the National Optical Astronomy Observatories, which are operated by the Association of Universities for Research in Astronomy, Inc., under cooperative agreement with the National Science Foundation.} routines in a PyRAF\footnote{PyRAF is a product of the Space Telescope Science Institute, which is operated by AURA for NASA.} environment. Nearby Landolt photometric standard fields SA-26, -29, and -32 (Landolt, \cite{landolt13}) were also observed on nights with appropriate photometric quality, and were used for determining standard magnitudes of a set of stars in the field of the target. These in turn were used to obtain standard magnitudes of the target itself for all dates. The data were corrected for interstellar reddening using the maps of Schlafly \& Finkbeiner (\cite{schlafly}).
Table\,\ref{tab:bvri-obslog} lists the $BVR_CI_C$ observations.


\begin{table*}
\centering
\caption{The log of $BVR_CI_C$ observations$^a$ of EPIC\,21175936.}
\label{tab:bvri-obslog}
\begin{tabular}{ccccccccc}
\hline\hline \noalign{\smallskip} 
{HJD} & $B$ & $\pm$ & $V$ & $\pm$ & $R_C$ & $\pm$ & $I_C$ & $\pm$\\
$[2\,400\,000+]$  & mag. & & mag. & & mag. & & mag. & \\ 
\hline
\noalign{\smallskip}\hline           
58190.4475  & 12.574  & 0.010 & 11.646  & 0.005 & 11.100  & 0.004 &10.444 & 0.003  \\ 
58191.3205  & 12.567  & 0.010 & 11.654  & 0.014 & 11.108  & 0.016 &10.442 & 0.028  \\   
58211.3385  & 12.596  & 0.023 & 11.635  & 0.008 & 11.088  & 0.002 & 10.438& 0.003   \\   
58213.4545  & 12.436  & 0.025 & 11.561  & 0.013 & 11.022  & 0.008 &10.383 & 0.014  \\   
58217.3445  & 12.407  & 0.006 & 11.492  & 0.004 & 10.964  & 0.008 &10.333 & 0.009  \\   
58220.3445  & 12.472  & 0.017 & 11.555  & 0.005 & 11.023  & 0.007 &10.382 & 0.009  \\   
58221.3395  & 12.496  & 0.015 & 11.583  & 0.008 & 11.041  & 0.010 &10.412 & 0.027  \\   
58222.3345  & 12.525  & 0.006 & 11.609  & 0.001 & 11.068  & 0.004 &10.415 & 0.011  \\   
58227.3405  & 12.553  & 0.011 & 11.628  & 0.003 & 11.090  & 0.006 & 10.434& 0.013   \\   
58228.3265  & 12.524  & 0.024 & 11.614  & 0.013 & 11.068  & 0.004 &10.419 & 0.035  \\   
58229.3295  & 12.514  & 0.006 & 11.589  & 0.003 & 11.048  & 0.005 &10.401 & 0.005  \\   
58230.3215  & 12.501  & 0.009 & 11.554  & 0.008 & 11.023  & 0.008 &10.385 & 0.006  \\   
58232.3425  & 12.425  & 0.023 & 11.534  & 0.003 & 11.017  & 0.025 &10.363 & 0.015  \\   
58233.3445  & 12.411  & 0.050 & 11.505  & 0.001 & 10.978  & 0.005 &10.380 & 0.015  \\   
58234.3325  & 12.443  & 0.018 & 11.532  & 0.009 & 10.987  & 0.001 &10.372 & 0.010  \\   
58236.3095  & 12.464  & 0.017 & 11.533  & 0.010 & 11.000  & 0.001 &10.367 & 0.001  \\   
58246.3395  & 12.608  & 0.004 & 11.661  & 0.005 & 11.110  & 0.005 &10.452 & 0.004  \\   
58247.3305  & 12.562  & 0.045 & 11.632  & 0.008 & 11.083  & 0.013 & 10.454& 0.020   \\   
58250.3355  & 12.430  & 0.006 & 11.535  & 0.009 & 11.004  & 0.009 &10.360 & 0.017  \\   
58251.3505  & 12.367  & 0.018 & 11.487  & 0.019 & 10.968  & 0.002 &10.339 & 0.016  \\   
\noalign{\smallskip}\hline                  
\end{tabular}

{\bf Notes.} $^a$Carried out at the Baja Observatory of the University of Szeged.
 
\end{table*}

\subsection{K2 Observations}

\begin{figure}[t]  
   \centering
      \includegraphics[width=0.98 \columnwidth]{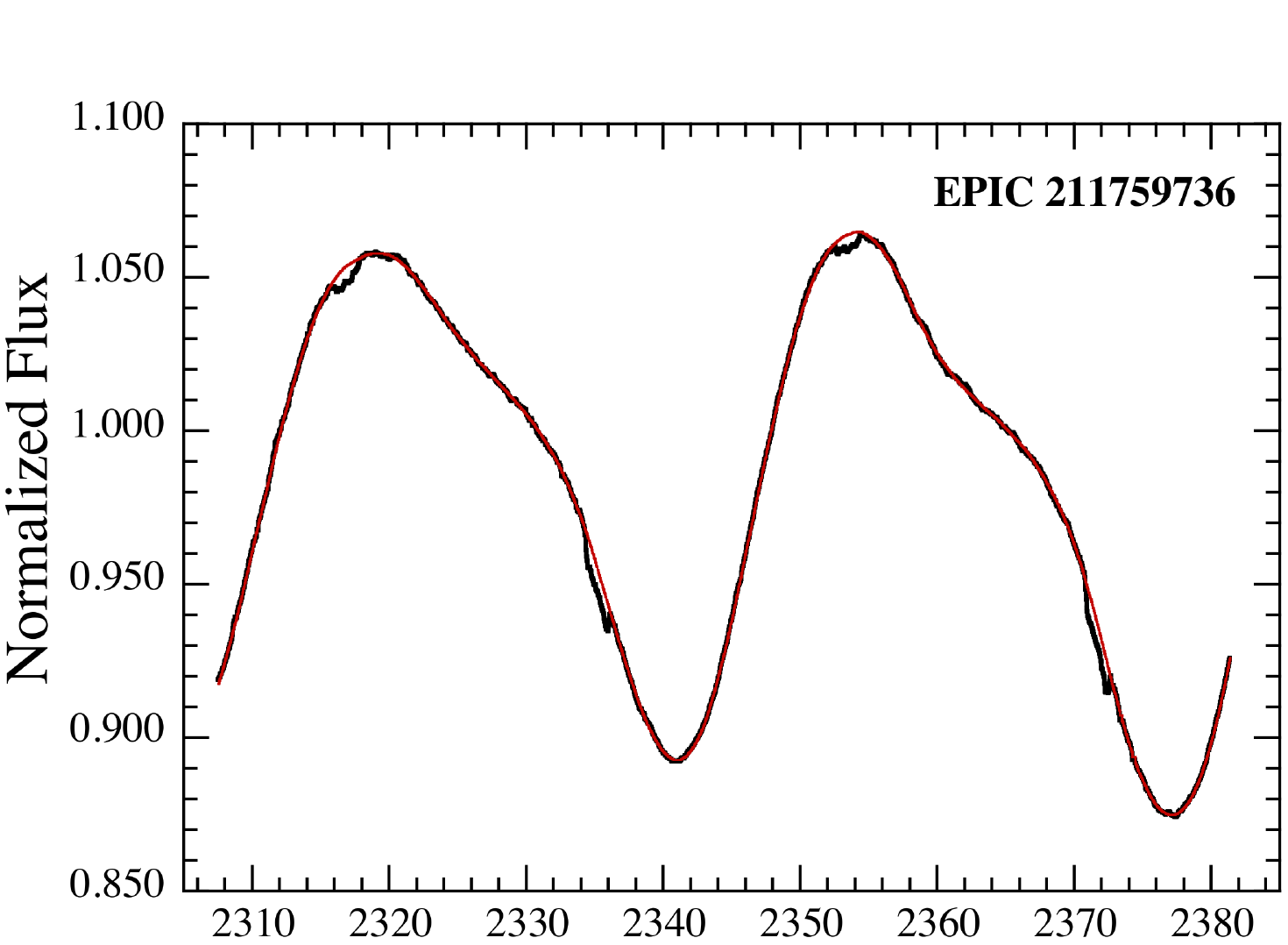} \\    
      \includegraphics[width=0.982 \columnwidth]{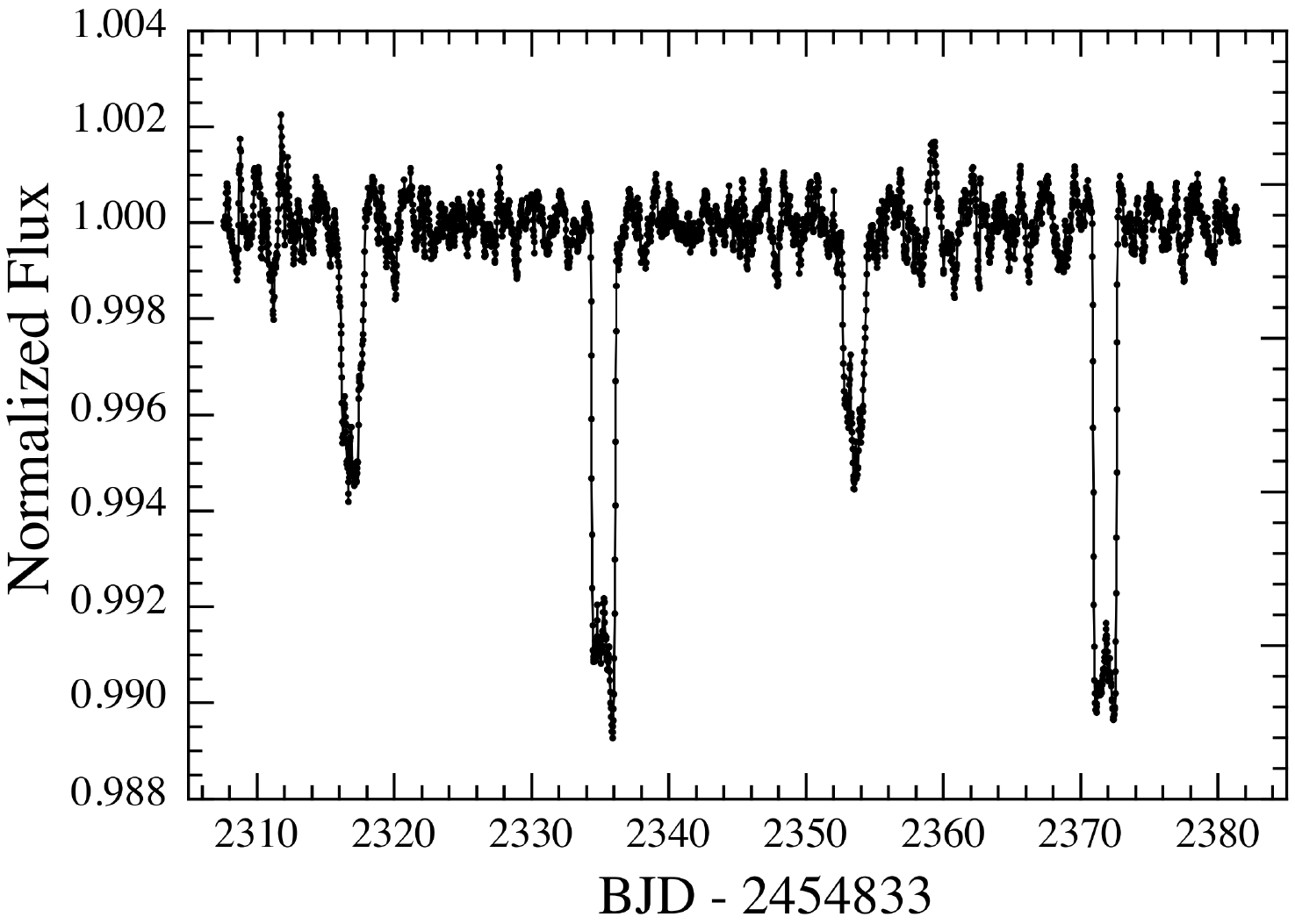}     
      \caption{K2 light curve of EPIC 211759736 spanning 75 days.  {\em Top panel:} The black curve is the raw light curve, while the thin red curve is a spline smoothed version.  {\em Bottom panel:}  The difference between the black and red curves in the top panel, showing more clearly both the primary and secondary eclipses.  Note the difference in vertical scale by a factor of $\sim$15. The eclipse depths are at most 5\% of the full amplitude ($\approx0.2$ mag) of the light variations.}
   \label{fig:K2}
 \end{figure}  

After the {\em Kepler} main mission, the {\em Kepler} spacecraft was re-purposed to observe a set of fields along the ecliptic plane. Each {\em Kepler} K2 campaign typically monitors some 25,000 stars in a given field for about 80 days (Howell et al.~2014), and a similar precision to that of the original {\em Kepler} mission is often achieved (see, e.g., Vanderburg et al.~\cite{vanderburg16}). EPIC 211759736 was observed during Campaign 5 (`C5') from April 28, 2015 to July 10, 2015.  The light curves were extracted from the {\em Kepler} pipeline calibrated target pixel files from the Mikulski Archive for Space Telescopes\footnote{MAST; https://archive.stsci.edu/}. The data were corrected for the K2 spacecraft-motion induced systematics following the approach described in Vanderburg \& Johnson (\cite{vanderburg14}) and Vanderburg et al.~(\cite{vanderburg16}). We also utilized the {\em raw data} of the very recent 2018 K2 {\em Kepler} observations of the star (Campaign 18; `C18').

In addition to systematic searches for periodic events, e.g., planetary transits, binary eclipses, and stellar pulsations, a number of Citizen Scientist groups visually inspect all the light curves by eye in search of aperiodic phenomena and/or events of an unusual nature that might well escape the systematic searches.  Such was the case here when two of us (T.\,J. and D.\,L.), using {\tt LcTools} (Kipping et al.~2015), found two pairs of shallow eclipses in the light curve of the highly modulated giant star EPIC~211759736.  

This K2 light curve for EPIC~211759736 is shown in Fig.~\ref{fig:K2}.  The top panel is the raw K2 light curve along with a spline fit to indicate the smoothly varying starspot modulations.  The bottom panel shows the difference between the data and the fit, thereby clearly revealing the primary and secondary eclipses.  The deeper eclipses are the giant passing in front of the higher-temperature, i.e., higher surface brightness, but much smaller, secondary star.  The more shallow eclipses, by contrast, occur when the smaller secondary star passes in front of the cooler giant and traces out the limb-darkened profile of the giant.

\subsection{Spectroscopic data}
\label{sec:TRES}

We observed EP211759736 with the Tillinghast Reflector Echelle Spectrograph (TRES; F\H ur\'esz et al. \cite{TRES}) on the 1.5 m Tillinghast Reflector telescope at the Fred Lawrence Whipple Observatory (FLWO) on Mt. Hopkins, Arizona. TRES has a spectral range of 3900-9100 Angstroms and a resolving power of R $\simeq$ 44,000. The spectra were reduced and extracted as described in Buchhave et al. (\cite{buch1}).

We obtained 10 radial velocity observations between UT 2018 February 04 and 2018 March 22. The spectra had an average signal-to-noise per resolution element (SNRe) of $\sim$30  at the peak continuum near the Mg b triplet at 519 nm with exposure times averaging 1200 seconds. A multi-order velocity analysis was performed by cross-correlating the spectra, order by order, against the observation with the strongest SNRe as a template. Twenty-two orders were used, excluding low S/N orders in the blue part of the spectrum and some red orders with telluric fringing.

The {\tt Stellar Parameter Classification} ({\tt SPC}; Buchhave et al. \cite{buch2}) tool was used to derive the stellar parameters of the giant. {\tt SPC} cross correlates the observed spectra against a library of synthetic spectra based on Kurucz model atmospheres (Kurucz et al. \cite{kurucz}). We calculated the weighted average of the parameters taking into account the cross correlation function peak height. Our results are fully consistent with those given in the EPIC input catalog (Huber et al. \cite{huberetal}).


\begin{table}[t]
\caption{Radial velocity measurements of EPIC~211759736}             
\label{radvel}      
\centering                          
\begin{tabular}{l l l l}       
\hline\hline \noalign{\smallskip} 
BJD & rad. vel. & error & $T_{\rm eff}$ \\
        & km/sec  & km/sec         & K   \\
\hline
2458153.846186  & 26.340     &  0.058  & 4668.06 \\
2458156.728078  & 41.115     &  0.055  & 4734.32 \\
2458170.820695  & 39.937     &  0.080  & 4706.88 \\
2458173.834510  & 22.483     &  0.098  & 4600.17 \\
2458179.747478  & 0.0000     &  0.080  & 4785.54 \\
2458181.614651  & $-$0.441  &  0.059  & 4773.20 \\
2458184.830288  & 4.7870     &  0.134  & 4702.90 \\
2458186.698832  & 10.669     &  0.086  & 4748.23 \\
2458189.707202  & 22.973     &  0.080  & 4708.55 \\
2458199.682131  & 60.673     &  0.072  & 4754.30 \\
\noalign{\smallskip}\hline                  
\end{tabular}
\end{table}

\section{Results}

\subsection{Long-term variations?}

\begin{figure}[tbp]
   \centering
      \includegraphics[width=8cm]{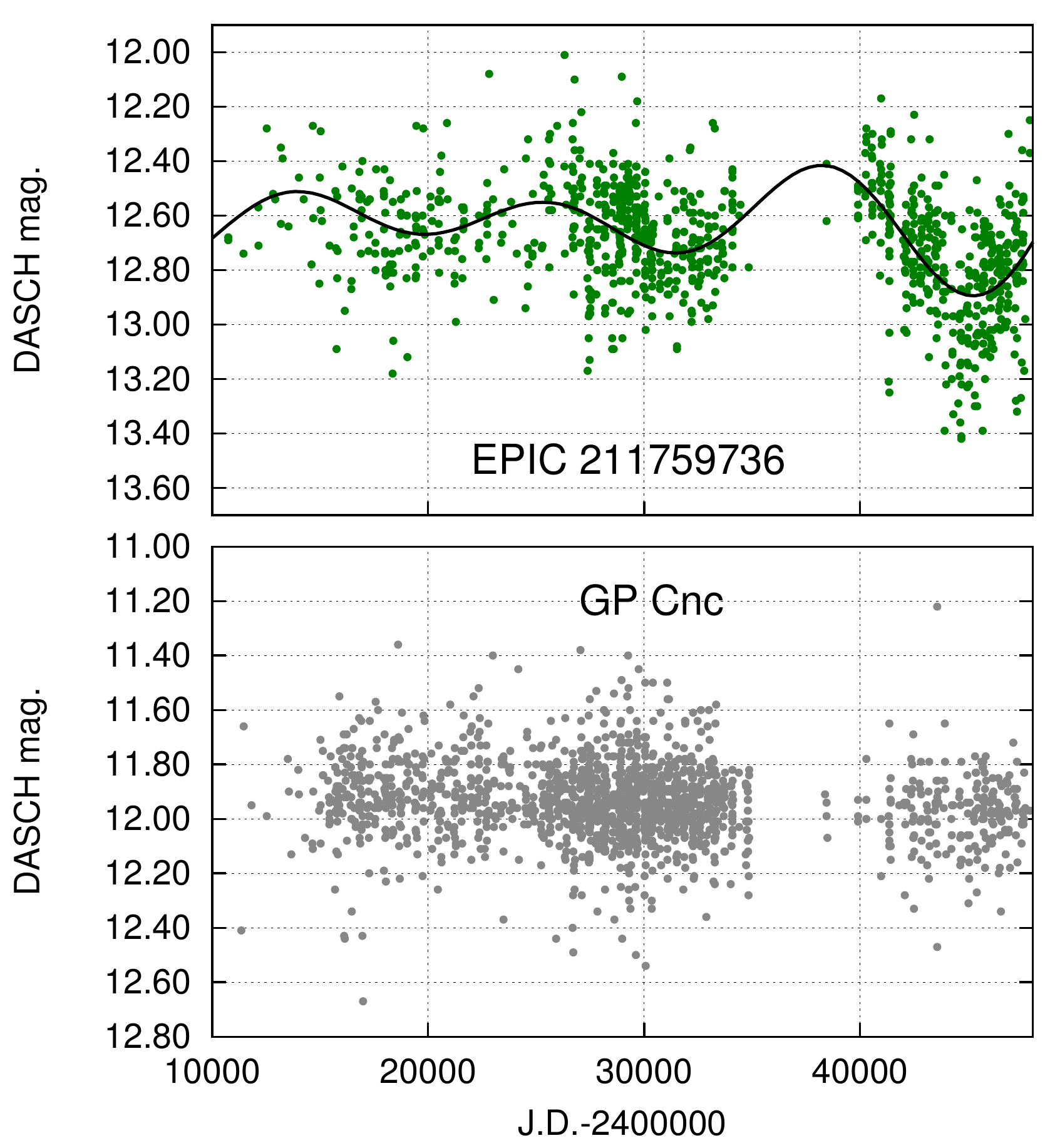}
   \caption{{\em Top:} Long-term photometric variations based on scanned photographic plates of EPIC~211759736 from the DASCH database.  Data are fitted with the possible decades-long cyclic changes of the star.  {\em Bottom}: For comparison, DASCH magnitudes of the nearby $\delta$~Sct-type star GP~Cnc, which is not expected to change its mean brightness. The two light curves are plotted to the same magnitude scale.}
   \label{dasch}
 \end{figure}

The long-term ($\sim$100 years) photometric history of EPIC~211759736 from the DASCH archival data (Grindlay et al.~\cite{grindlayetal}) is shown in Figure \ref{dasch}.  The historic photometry record shows hints of long-term changes possibly due to variations in general spottedness, i.e., an activity cycle.  This can be seen in Fig.~\ref{dasch} where a change on the timescale of a few decades ($\approx$40-yr) is indicated by the solid curve  drawn by the beating of two close periods. Another, weaker signal with a possible 7-8 years cycle time is also found. 
For comparison, the bottom panel of Fig.~\ref{dasch} shows the DASCH photometry of the $\delta$~Sct-type star GP~Cnc, with low amplitude light variation (below 0.1 mag.; Wetterer et al. \cite{ibvs}) and without long-term changes in its mean magnitude. GP~Cnc is very close to EPIC~211759736 on the sky, which rules out systematics in the long-term photographic-plate records of our object.  

\subsection{Rotational periods and differential rotation}

We extracted rotational periods from the data with the multiple-frequency analysis tool {\tt MuFrAn} (Csubry \& Koll\'ath \cite{mufran1}). 
The rotational period of the star is fairly long (about 10 rotations/year), therefore one ground-based observing season covers only a few (6-8) rotations.  Thus, more than one season of observations is advisable to obtain a well-determined rotation period. But the period is expected to change due to the anticipated differential rotation and the possible appearance-dissolution and movement of the spots. Therefore, time-spans longer than 2~years should not be used  to derive rotational periods, since this is the timescale on which the light variation is stable, i.e., the phases of the extrema of the light curve do not change appreciably. 

\begin{figure}[tbp]
   \centering
      \includegraphics[width=9cm]{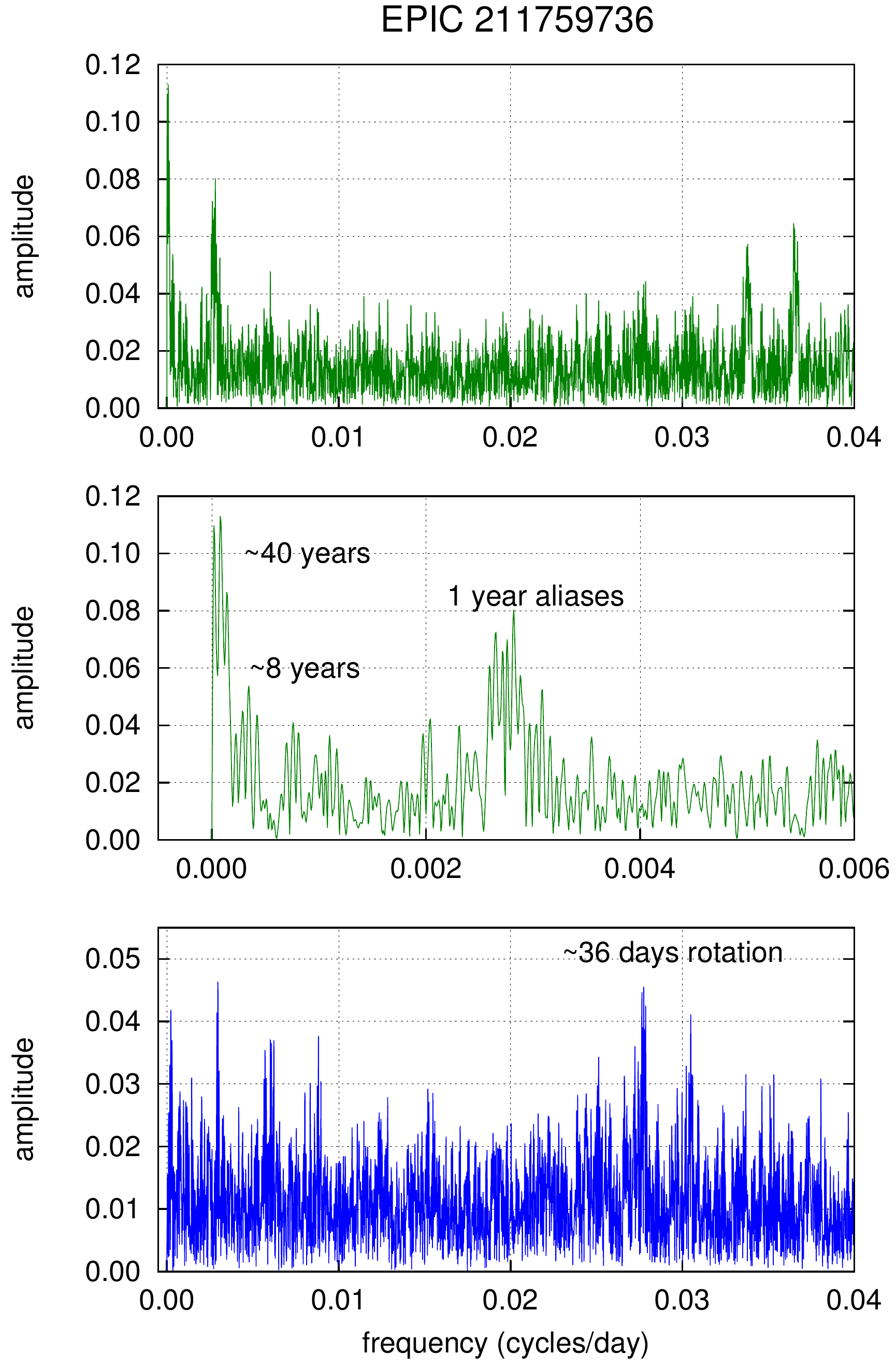}
   \caption{Fourier amplitude spectra of EPIC~211759736 from the DASCH dataset. {\em Top:} Fourier amplitude spectrum. {\em Middle:} Detail of the long-period range of the amplitude spectrum. {\em Bottom:} Weak signals around the 36-d rotational period after removing the long-term trends from the data.}
   \label{dasch_sp}
 \end{figure}

The Fourier transform (`FT') of the 100 years of DASCH photometric data for EPIC 211759736 is shown in Fig.~\ref{dasch_sp}.  The top two panels show the FT down to periods as short as 25 days, and a zoom-in on the lower-frequency portion of the spectrum, respectively.  Not much evidence for the 36-day rotation period is seen.  However, after removing the long-term variation from the dataset, we find a group of weak peaks near periods around 36 days, indicating that the rotational signal of the giant is indeed present in the historical data.  See Figure \ref{dasch_sp}.

\begin{figure}[tbp]
   \centering
      \includegraphics[width=8.5cm]{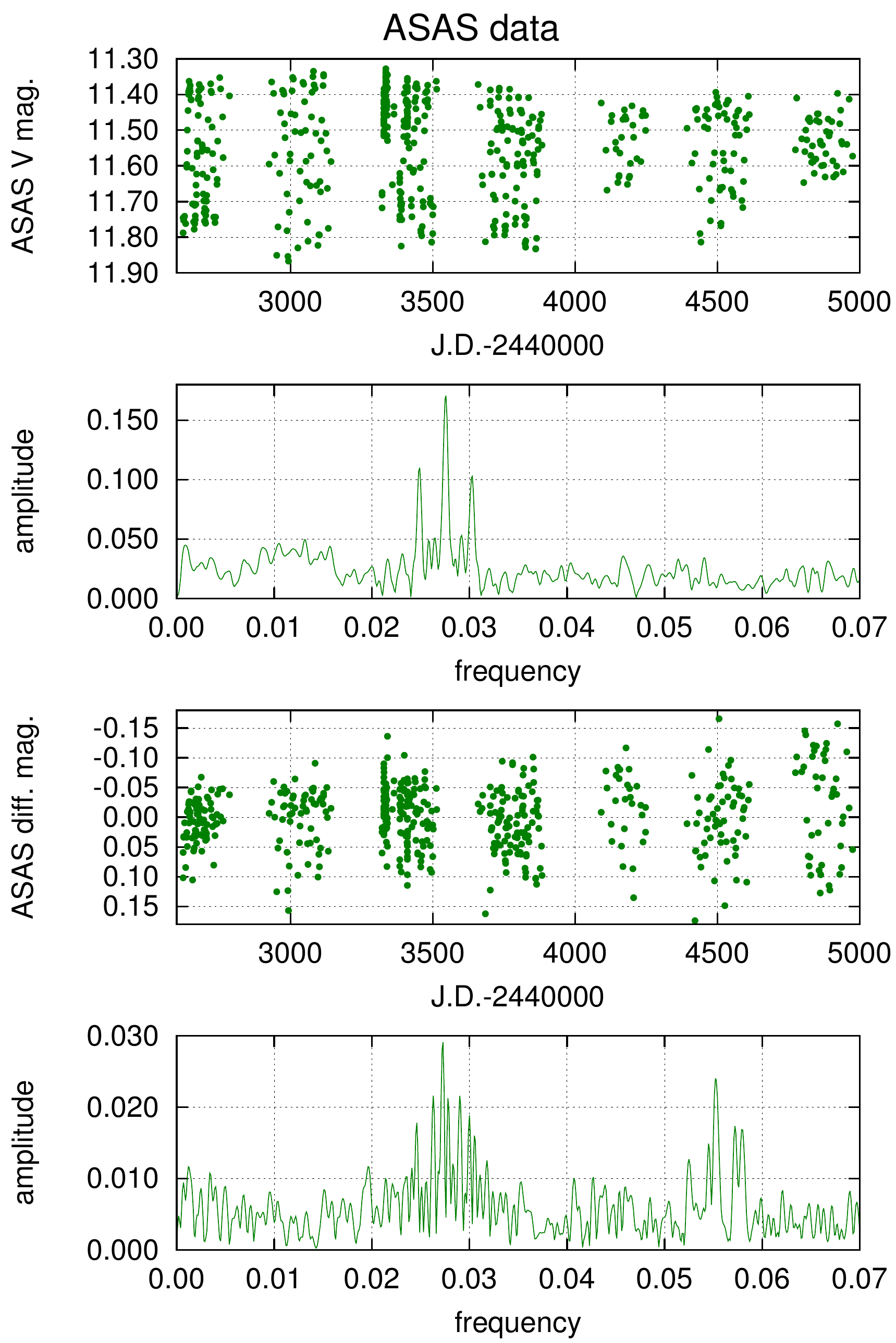}
   \caption{{\em Top two panels:} The 7-yr long ASAS dataset and its Fourier amplitude spectrum resulting a 36.27 days rotational period.  The two `satellite peaks' in the amplitude spectrum are the sidebands of the 1-year observational window function. {\em Bottom two panels:}  Data `cleaned' of the rotational period, and its amplitude spectrum. Apart from the remaining signals near the 36-d periods, half periods also show up significantly due to the non-sinusoidal light curves in the second half of the dataset.}
   \label{ASASfig}
 \end{figure}

We did a period search of the entire 7-yr long ASAS dataset (Pojmanski \cite{ASAS}).  The raw ASAS light curve is plotted in the top panel of Fig.~\ref{ASASfig}.  The rotational period for all the data is 36.27 days (second panel of the figure). After removing this signal, the difference light curve (third panel of Fig.~\ref{ASASfig}) shows an increased amplitude during the last three years relative to the previous ones, and the half period appears in the Fourier amplitude spectrum (fourth panel). This indicates that the light curve was sinusoidal during the first 4 years, but from 2006 onward the light curves became asymmetric and the half period appeared significantly with about 1/3 the amplitude of the fundamental period, suggesting two well separated active regions on the stellar surface. Additionally, in the case of both the ASAS and HAT surveys we used 1-2 seasons of data to derive independent periods for different epochs. We list the rotational periods in Table~\ref{periods} at different epochs found from the ASAS and HAT surveys. The 7th year of the ASAS data and the 1st one of the HAT data (2008-2009) cover nearly the same time interval, and the consistency of the results demonstrates the reliability of the derived periods. The resulting rotational periods of these overlapping observations are within their mutual uncertainties. 

\begin{figure}[tbp]
   \centering
      \includegraphics[width=9cm]{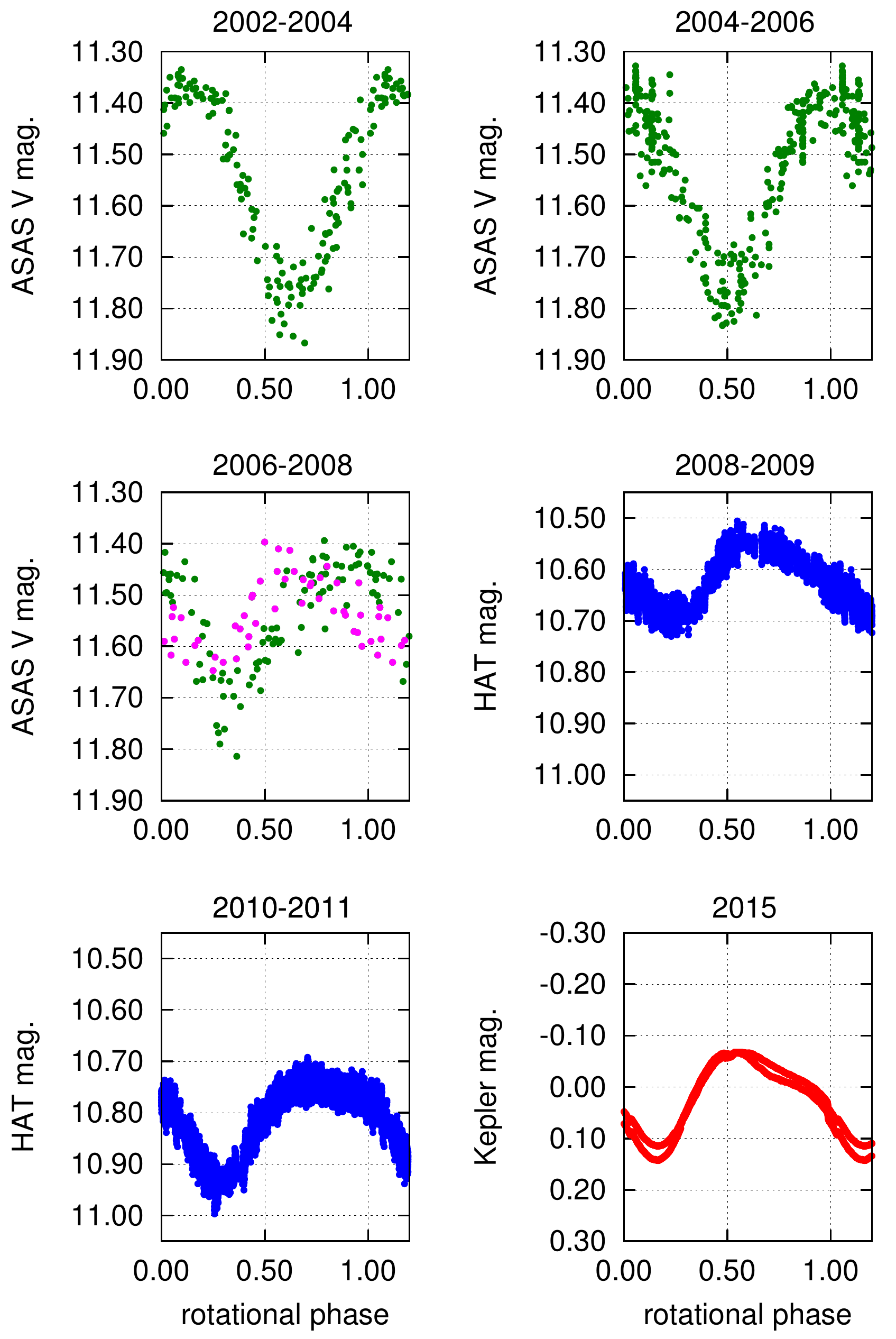}
   \caption{Light curves of EPIC~211759736 folded using the time of primary minimum and the orbital period from Table~\ref{specorbit}. ASAS data from 2008-2009 are overplotted with magenta points on the 2006-2008 ASAS light curve. Note, that the different flux levels are not real and merely reflect the different observational bandpasses and automated reduction processes (HAT). The deeper minimum is always between phases of about 0.1-0.6.}
   \label{lightcurves_orbital}
 \end{figure}

We plotted all the photometric data, folded with the orbital period (see footnote to Table~\ref{periods}), in Fig.~\ref{lightcurves_orbital}. 
This is the only reliable and completely coherent period in the system, and differs from the rotational period by only about 0.2 day, with the orbital period being longer.

\begin{table}[tbh]
\caption{Observed rotational periods of EPIC~211759736}             
\label{periods}      
\centering                          
\begin{tabular}{c c c c l}       
\hline\hline \noalign{\smallskip} 
survey & JD start - end & years & $P_{rot}{^1}$ & ampl.${^2}$ \\
\hline
ASAS  &  52623 - 53143 & 2002-2004 & 36.32  & 0.21  \\
ASAS  &  53322 - 53884 & 2004-2006 & 36.39  & 0.19 \\
ASAS  &  54092 - 54612 & 2006-2008 & 36.35  &  0.13 \\
            &                                &                   &18.11   &  0.04 \\
ASAS  & 54774 - 54975  & 2008-2009 & 36.30  & 0.08 \\
HAT    &  54792 - 54971 & 2008-2009 & 36.22  & 0.07 \\
           &                                &                   & 18.11   &  0.02 \\
HAT    &  55507 - 55674 & 2009-2010 & 36.40  & 0.09  \\
           &                                &                   & 18.19  & 0.03  \\
\noalign{\smallskip}\hline  
Average & 52623 - 55674 &  2002-2010 & $36.33 \pm 0.03$ & ... \\
\noalign{\smallskip}\hline 
                
\end{tabular}
\tablefoot{For comparison: $P_{\rm orb}$ = 36.522 days. ${^1}$The uncertainties in the spot-rotation periods are approximately 0.2 day, which means that periods off more than this amount result in noticeable changes in the folded light curve.  ${^2}$Fourier amplitude.} 
\end{table}

\subsection{Starspots and spot temperatures from $BVR_CI_C$ data}\label{5col}

There are only two active giant stars, $\zeta$~And and $\sigma$~Gem, which have already been directly mapped using interferometry and they exhibit large scale magnetic structures on their surfaces (see Roettenbacher et al., \cite{rachael1}, \cite{rachael2}). By contrast, the starlight we observe from EPIC 211759736 comes from a point source, and only approximate inferences can be made about the starspots' structure. We know from the Sun that the spots are not uniform (umbra-penumbra) and the larger active regions (in solar terminology `active nests') contain both dark (cool spots) and bright (hot faculae) areas. On other stars, very probably, we are observing a mixture of these regions with an averaged effective  temperature. In the following we use both the terms `spots' and `active regions', meaning the areas on the stellar surface where the activity is concentrated.

\begin{figure}[tbp]   
   \centering
      \includegraphics[width=0.99\columnwidth]{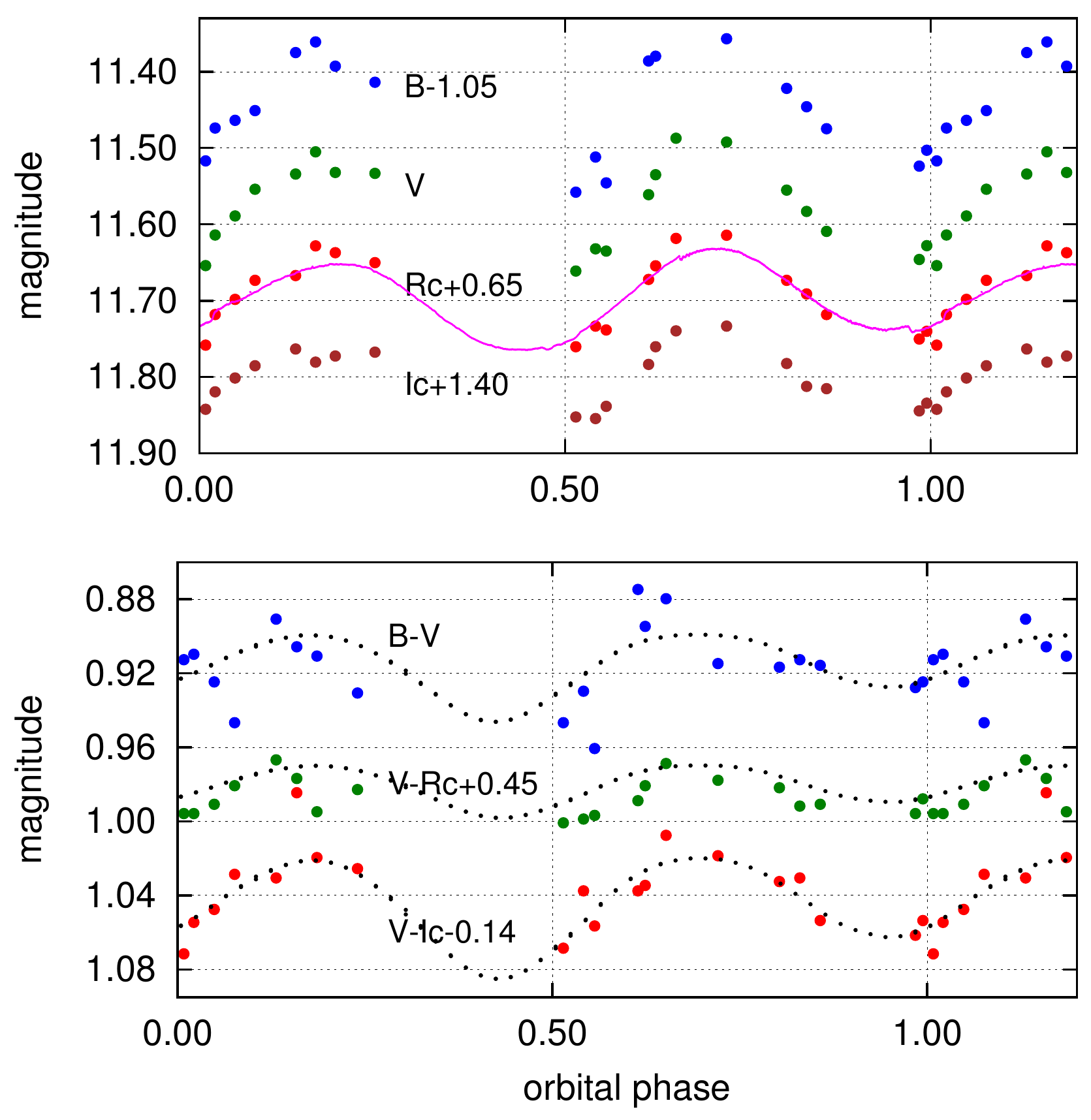}
           \caption{$BVR_CI_C$ light curves and color indices of EPIC~211759736 folded using the time of primary minimum and the orbital period from Table~\ref{specorbit}. At this epoch a clear double humped light curve is observed. The  magenta curve represents the nearly contemporaneous K2 {\it Kepler raw} data from C18.
           The three color indices are fitted with 3960~K spots (dots), see text for details.}
   \label{BVRI_data}
 \end{figure}

The shape of the photometric variations due to starspots in EPIC~211759736 (see Fig.~\ref{lightcurves_orbital}) changes relatively slowly. Between 2002 and 2007 the basic light curve shape was nearly sinusoidal and was probably caused by a single dominant region on the surface of the giant.  If there were more  than one region they were likely close to each other in longitude. From 2008 the light curves started to reveal two dominant, presumably detached spotted regions, manifested by non-sinusoidal light curves (see Fig.~\ref{lightcurves_orbital}).  By the time of the K2 observations in 2015, a small secondary maximum appeared showing that the spotted regions had moved farther away from each other. The double-humped light variation observed in $BVR_CI_C$ colors and by K2 during 2018, shown in Fig.~\ref{BVRI_data}, upper panel, clearly demonstrates two active regions on the stellar surface with the maximum possible longitude difference of about 180$^\circ$ between them. The dominant minimum of the light curve drifts slowly from phase 0.6 to phase 0.1 in Fig.~\ref{lightcurves_orbital} showing that the average rotational period of the star is indeed shorter than the orbital one. In 2018 the light curves have two minima of nearly equal depths and one of those is near phase 0.9, continuing the migration of a long-lasting (from 2002 to date) active region. The observed slow drift of the light curve minima is possibly due to the latitudinal differential rotation (see Table~\ref{periods}).

To model the spotted light variations in the $BVR_CI_C$ dataset we used our own software based on the analytical equation applied to spot modeling by Budding (\cite{budding}), assuming circular spots. Fixed parameters were the effective wavelengths of the filters used, linear limb darkening coefficients (Howarth, \cite{howarth}) taking into account the stellar parameters, and the unspotted brightness in all bandpasses.  We note that there is no way to deduce the true unspotted brightness of an active star showing the usual rotational and long-term light variations. On the other hand, one would expect a higher brightness level and smaller rotational variation with fewer and fewer spots. Therefore, we took the observed maximum magnitude as the unspotted reference brightness level. The rest of the activity is supposed to be distributed evenly and/or remain on the poles. Note that we have observed only two rotations of the star in a standard photometric system, and the archival and K2 data were obtained in different, non-standard bandpasses (except ASAS which has a $V$ magnitude bandpass), and therefore cannot be compared quantitatively.

We modeled in parallel the $B-V$, $V-R_C$, $V-I_C$ color observations, assuming that there are just two circular starspots. The parallel modeling of each of these colors results in spot coordinates (longitudes, latitudes), sizes, and a single spot temperature. For EPIC~211759736 we have a high rotational inclination angle (81.85$^\circ$), and thus the two stellar hemispheres are nearly interchangeable (supposing that the rotational axis is perpendicular to the orbit, the star is seen nearly edge-on), i.e., approximately invariant under inversion. Therefore, in the course of the modeling, the spot latitudes were kept fixed at the equator; generally, it is not possible to obtain reliable spot latitudes from photometric data. In this way we had 5 free parameters to fit for: two longitudes and sizes, and the spot temperature. 

The results show that in early 2018 there were two cool spots (active regions) on the stellar surface, one facing the secondary component and the other on the opposite side, covering altogether about 10\% of the total stellar surface, with a temperature of $3960 \pm 300$~K, or about 800~K below the surface temperature of 4750~K. The individual results of the spot temperature modeling from the three different color indices are as follows: $B-V: 3968 \pm 305$, $V-R_C: 3936 \pm 364$, and $V-I_C: 3978 \pm 215$.  Although the spot temperatures from the three color indices have substantial uncertainties, their values are remarkably close to each other. The resulting spot latitudes and sizes from the three color indices are within their mutual 1-{$\sigma$} uncertainties. The 4-color light curves and the fitted color indices are plotted in Fig.~\ref{BVRI_data}. (We note that experiments allowing the spots' latitudes to also be free parameters gave essentially the same results for the longitudes, sizes and temperature, but with much higher uncertainty due to the error propagation.)

By contrast, the average stellar temperatures from the color indices are as follows: $B-V : 5003 \pm 48, V-Rc: 4759 \pm 39, V-Ic: 4456 \pm 36$, where the errors are rms values.  By comparison, the photospheric temperature from the TRES spectra is $4734 \pm 93$~K (rms). The three different results from the color indices show the presence of bright, hotter faculae (from $B-V$) and cool spots (from $V-I_C$), and a combination of these is reflected in the spot temperatures.  Again, active regions are generally presumed to consist of both hotter and cooler regions than the surrounding quiet photosphere.

\section{Binary modeling and spots from K2 data}
\label{binary_model}

\begin{figure}[tbp]
   \centering
      \includegraphics[width=8cm]{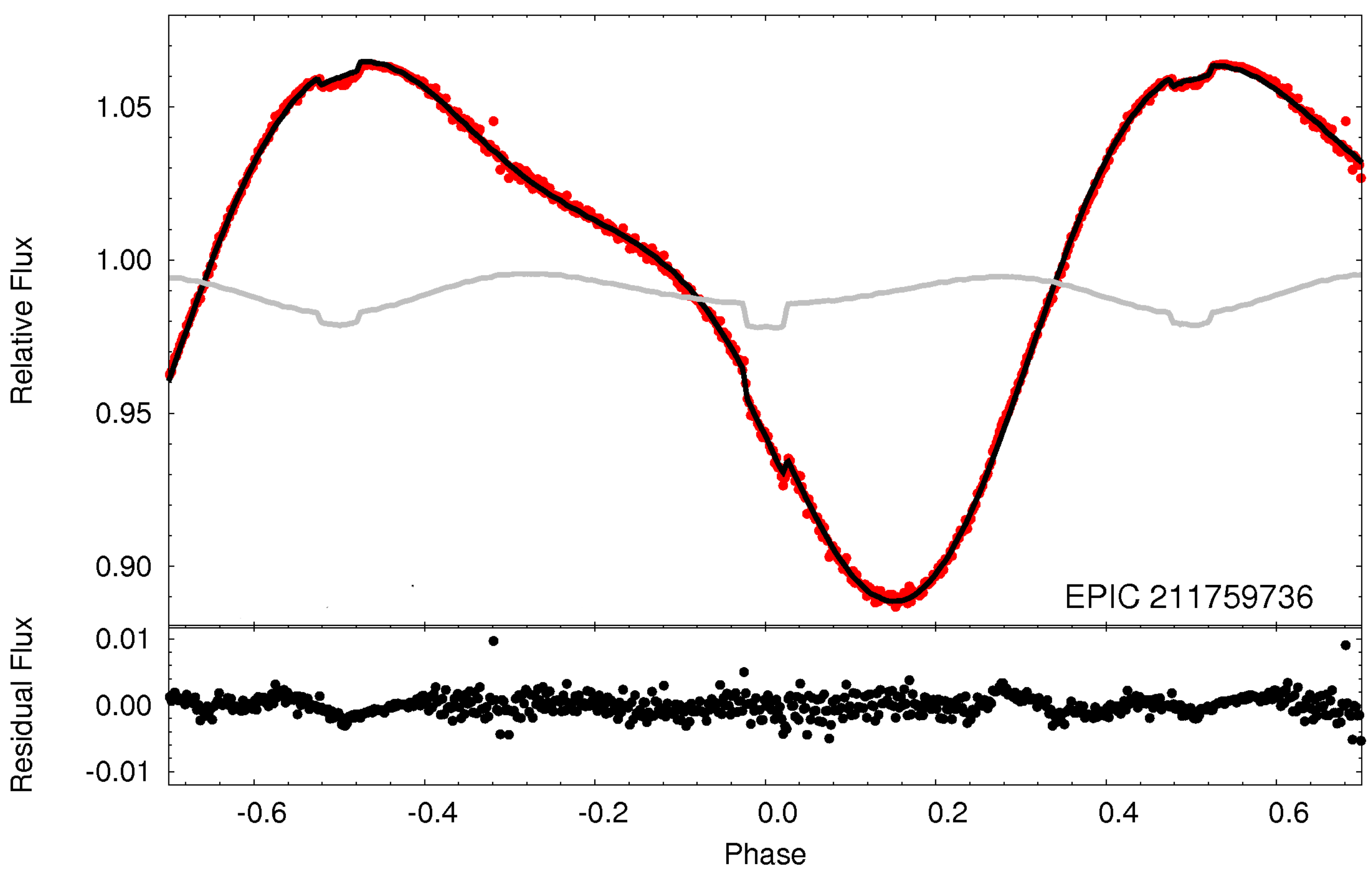}
   \caption{The orbital-phase folded $K2$ light curve of EPIC\,211759736 from C5 (red points), together with the pure EB model (grey) and the sum of the EB model and the mathematically described photospheric variations (black, see text for details and Eqn.~(\ref{eqn:spots}) in particular). The lower panel shows the residuals after subtracting the combined light curve model from the data.}
   \label{fig:foldedlightcurve}
 \end{figure}

The accurate, two orbital-period-long $K2$ photometry made it possible to disentangle the brightness variations arising from the binarity and the spottedness.  In order to do this, we first made an orbital phase-folded, binned mean light curve from the $K2$ data from C5 (see Fig.~\ref{fig:foldedlightcurve}). Then we carried out a simultaneous analysis of this folded $K2$ light curve and the RV curve with our MCMC-based light curve emulator code {\tt Lightcurvefactory} (Borkovits~et~al.,\,\cite{Borko13}, Rappaport~et~al.,\,\cite{Rappaport17}). Similar to the method described in Borkovits~et~al.\,(\cite{borkovitsetal18}), the light curve variations arising from stellar spots instead of the binarity, are simply modeled  mathematically by a harmonic function of the form:
\begin{equation}
\Delta\mathcal{L}=\sum_{i=1}^4a_i\sin(2\pi f_it)+b_i\cos(2\pi f_it),
\label{eqn:spots}
\end{equation}
where the four frequencies ($f_1=0.027367\,\mathrm{d}^{-1}$, $f_2=0.054747\,\mathrm{d}^{-1}$, $f_3=0.078965\,\mathrm{d}^{-1}$, $f_4=0.134952\,\mathrm{d}^{-1}$) represent the four highest peaks in the Fourier spectrum of the $K2$ light curve from C5.  The coefficients $a_i$ and $b_i$ are calculated via a linear least-squares fit.  This function is applied to the residual light curve formed by subtracting the pure eclipsing binary model from the observed light curve, at each step in the MCMC process. Then, this mathematical model of the residual light curve is added to the binary model light curve, and the actual $\chi^2$ value is calculated for this mixed model light curve.

In order to obtain the preliminary, unspotted binary model, the following nine parameters were adjusted:
\begin{itemize}
\item[(i)]{5 orbital parameters: $P_\mathrm{orb}$, eccentricity $e$, argument of periastron $\omega$, inclination $i$, time of periastron passage $\tau$;}
\item[(ii)]{2 RV-curve-related parameters: systemic velocity $\gamma$ and spectroscopic mass function $f(m_2)$;}
\item[(iii)]{2 light curve related parameters: duration of the transit of the main-sequence secondary component ($\Delta t$), and the temperature ratio ($T_2/T_1$) of the two components.}
\end{itemize}

\begin{figure}[tbp]
   \centering
      \includegraphics[width=8.5cm]{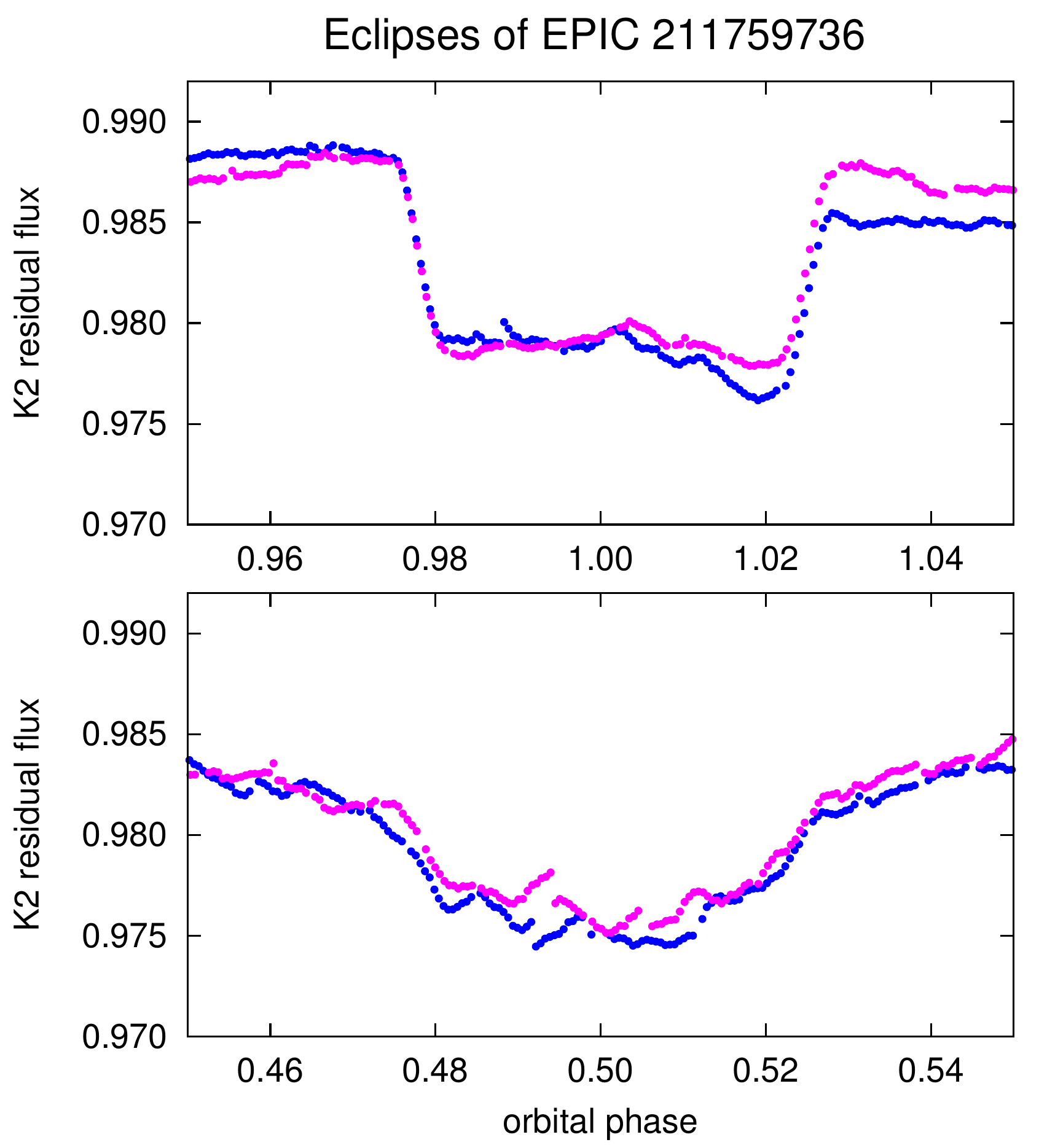}
   \caption{The two eclipses of EPIC~211759736 from C5, from the first (blue) and second (magenta) rotations with the rotational modulation removed. Phases were calculated using the time of primary minimum and the orbital period from Table~\ref{specorbit}. {\it Top:} primary minimum (giant eclipses secondary), {\it bottom:} secondary minimum (secondary star transits the giant). }
   \label{eclipses}
 \end{figure}

\begin{figure}[tbp]
   \centering
      \includegraphics[width=9 cm]{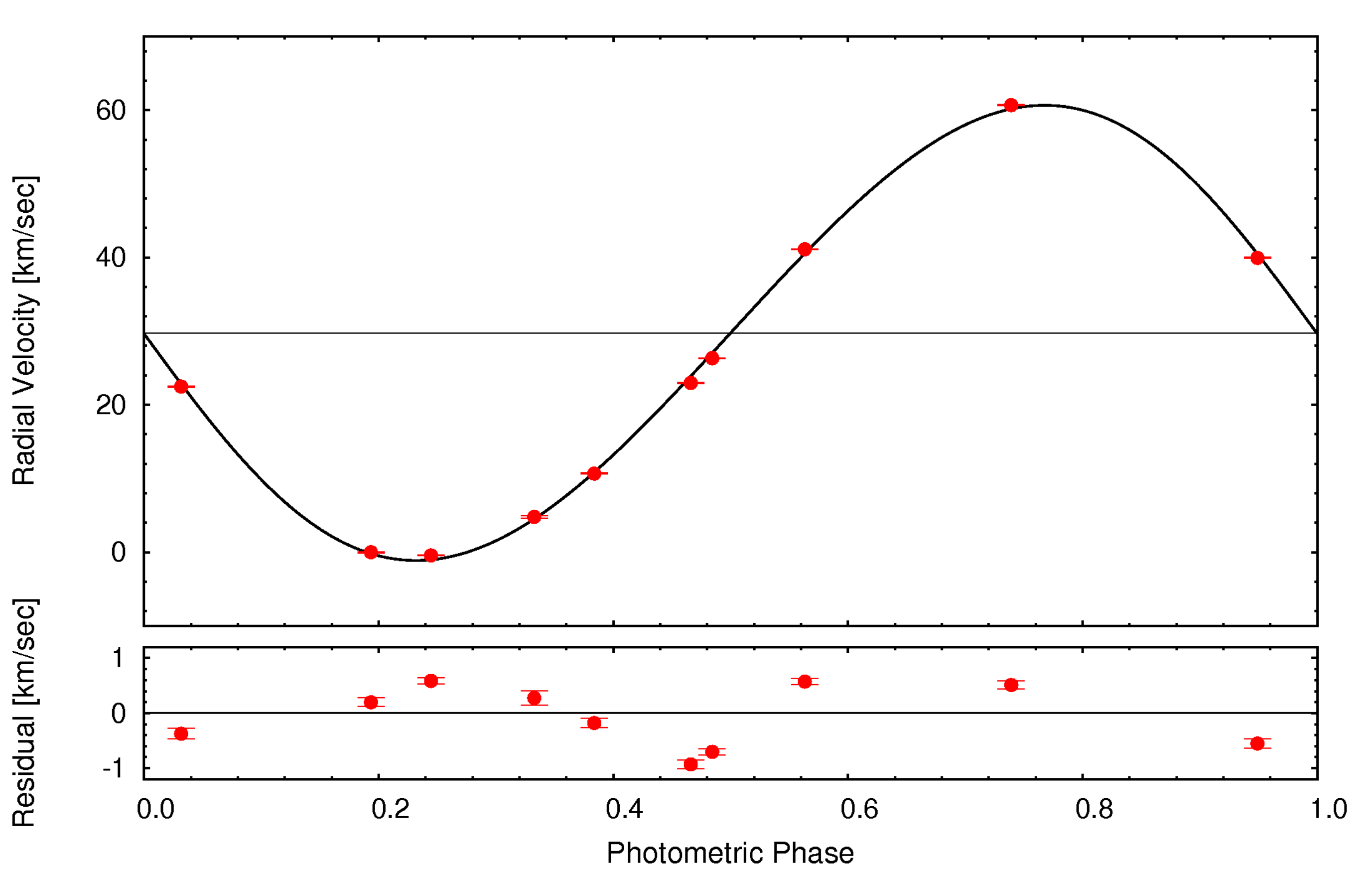}
   \caption{Radial velocity curve for EPIC~211759736 obtained with the Tillinghast Reflector Echelle Spectrograph (TRES).  See Sect.~\ref{sec:TRES}.}
   \label{RVfit}
 \end{figure}

The temperature of the giant component was taken from our spectroscopic measurements (via template fitting; see Sect.\ref{sec:TRES}). The resulting temperature of $4734\pm93~K$ was rounded to 4750~K (Table \ref{params}). Furthermore, regarding the secondary component, we assumed that it is an unevolved MS star: note the difference between the rounded limb-darkened profile of the secondary eclipse and the sharp ingress and egress of the primary eclipse in Fig.~\ref{eclipses}.  In keeping with this assumption, the mass and the radius of the secondary were calculated internally in the fitting code from its effective temperature via the use of the main-sequence $T(M)$ and $R(M)$ relations of Tout~et~al.\,(\cite{Tout}).
The parameters we obtained from this fit are listed in Tables\,\ref{specorbit} and \ref{params}, while the RV solution is plotted in Fig.\,\ref{RVfit}.  These results are in good agreement with the {\it Gaia} parameters listed for this system ({\it Gaia} Collaboration et al., \cite{GAIA1}, \cite{GAIA2})  especially the temperature which is essentially the same, but the radius and luminosity values are also within their mutual 2~$\sigma$ and 3~$\sigma$ error bars (see the Introduction).

\begin{figure}[tbp]
   \centering
      \includegraphics[width=8.7cm]{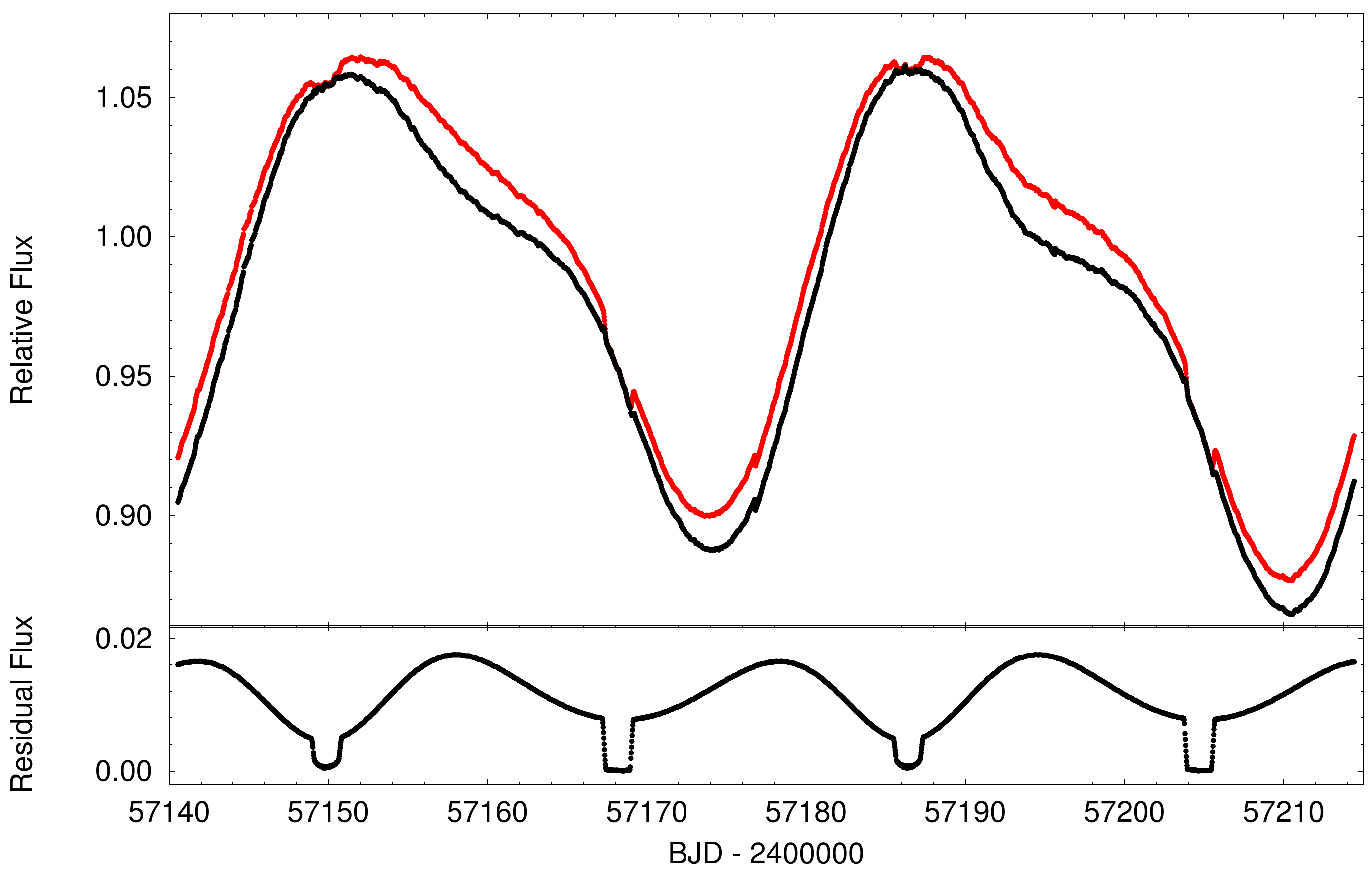}
   \caption{The original, preprocessed $K2$ light curve (red dots) together with the spottedness-only light curve (black curve, upper panel), obtained by the removal of the binarity-produced light variations, i.e. the eclipses, ellipsoidal light variation and Doppler boosting effect (lower panel).}
   \label{fig:disentangledK2lightcurve}
 \end{figure}

Once we have the basic orbital and stellar parameters determined, we can examine the starspots that were present on EPIC~211759736 during the K2 observations.  We proceed by subtracting off the  pure EB light curve from the original $K2$ light curve (see Fig.\,\ref{fig:disentangledK2lightcurve}).  


  \begin{figure}[tbp]
   \centering
      \includegraphics[width=8.9 cm]{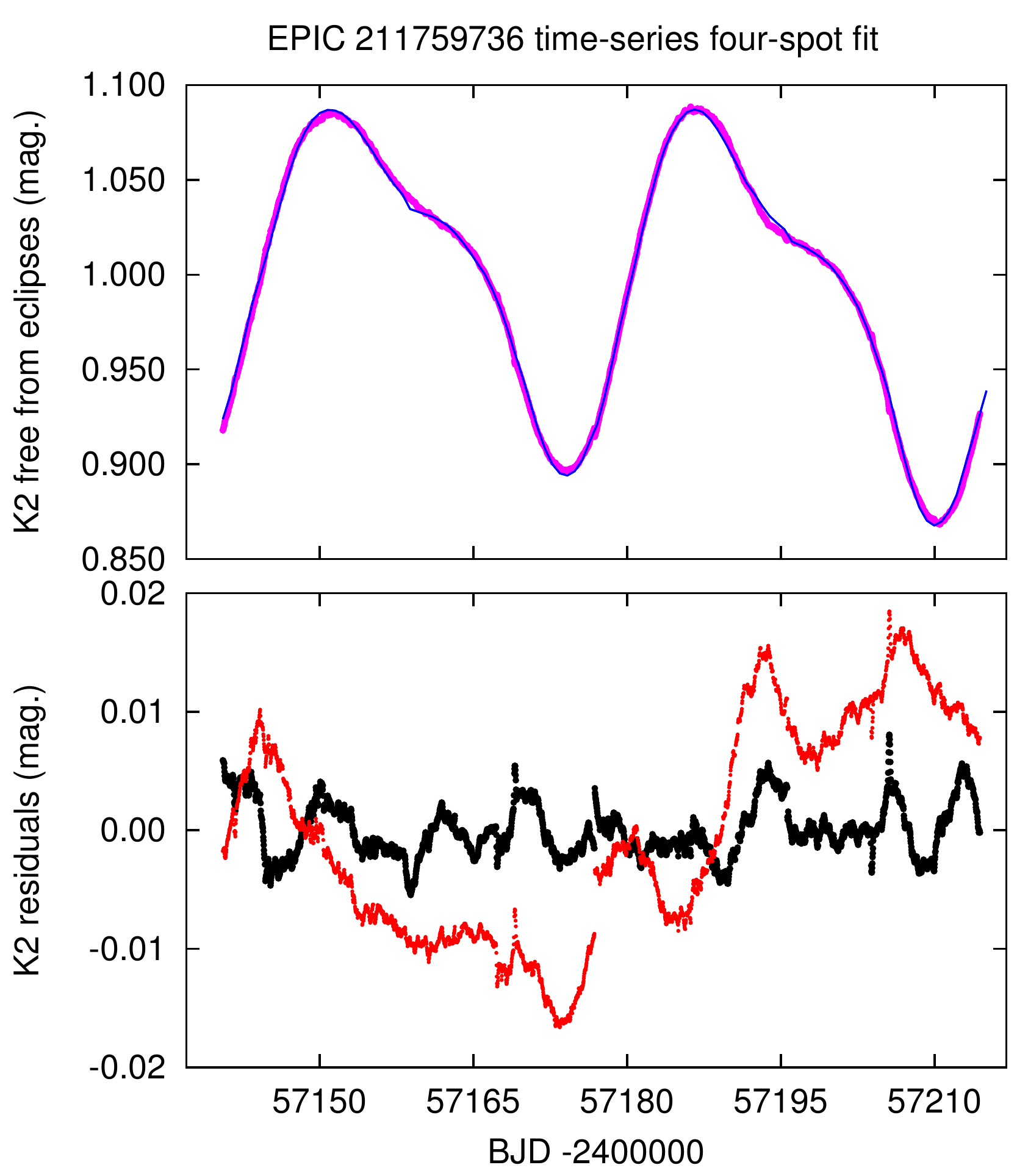}   
      \caption{Time-series spot modeling of the $K2$ data free from the eclipses, ellipsoidal light variations, and any other effects of binarity, i.e., showing the spot modulations only. In the upper panel the $K2$ light curve is plotted in magenta, together with the almost identical fit from our 4-spot model (blue line). The residuals from the 4-spot model are plotted in the lower panel (black curve), while the deviations of the 2-spot model from the data is plotted in red.}
   \label{spots_timeser}
 \end{figure}  
 
In the course of the modeling we took the spot temperature to be that derived from the $BVR_CI_C$ data (i.e., 3960~K, see Sect. \ref{5col}), since the \emph{Kepler} data have only one bandpass ($\sim$4500 to 8500 $\AA$), i.e., with basically no colour information. First, we assumed two active regions, but since the precision of the {\em Kepler} photometry is very high, we set the spots' latitudes, longitudes, and sizes as free parameters to be fitted. As another experiment we put 4 spots on the stellar surface with fixed latitudes: two spots were put at $20^\circ$ above the equator, and two spots at $-20^\circ$, below the equator, this way allowing for 8 free parameters. The fit of the time-series analysis from the 4-spot model is shown in Fig.~\ref{spots_timeser}. The small temporal change of the light curve over the two subsequent rotations can be well understood with small motions and size variations of the assumed circular spots (active regions), probably reflecting the emergence and decay of smaller spots within the assumed circular active regions. The goodness of the fit is seen in the lower panel of Fig.~\ref{spots_timeser} for the 2-spot model (in red color) and the 4-spot model (in black) as well.

The formal errors in the fitted spot longitudes for the $K2$ data are always $\lesssim 1^\circ$, while errors in the spot latitudes (if set as free parameters in the 2-spot model) are between $1^\circ$ and $2^\circ$, and the radii are accurate to a few tenths of a degree. As a comparison, in the case of the ground-based data, the errors in the spot longitudes are about $1^\circ - 2^\circ$, while the spot latitudes are indefinite. Furthermore, using ground-based observations, we find that the spot radii are accurate only to $3^\circ-4^\circ$ since the fit is not ideal due to the fixed spot latitudes and the much larger observational errors. We note that these errors should be regarded as internal errors of the method, i.e., they do not reflect true uncertainties in the actual spots since those are not simple circular or steady features. An early paper by K\H ov\'ari \& Bartus (\cite{test}) gives some good insight into the drawbacks of photometric starspot modeling.

\begin{figure}[tbp]
   \centering
      \includegraphics[width=8 cm]{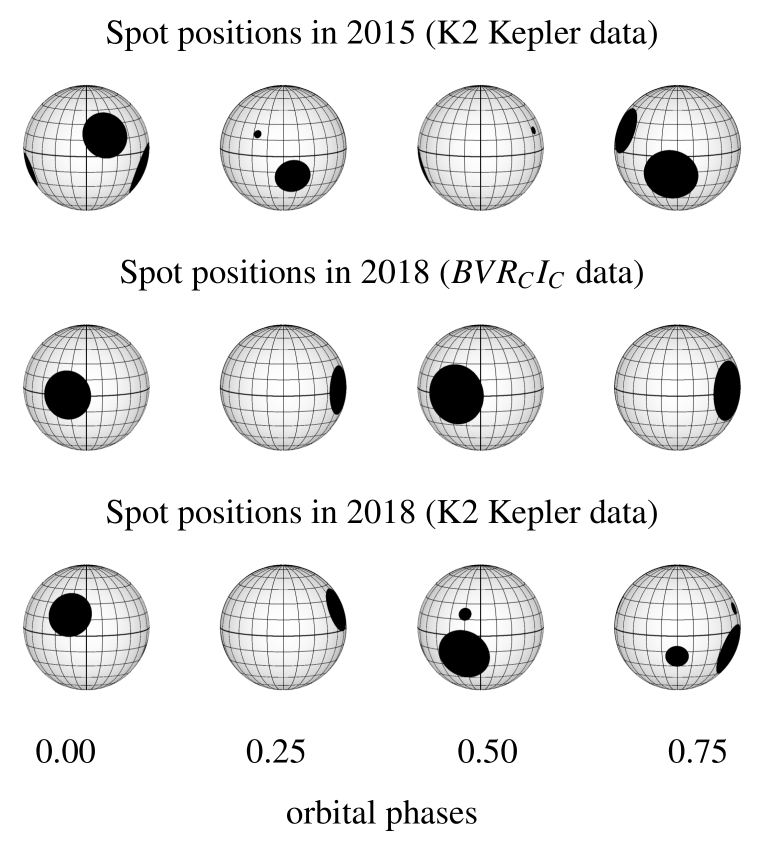}
    \caption{The positions of the active regions in 2015 from our 4-spot model fit to the $K2$ C5 data (top row)
and in 2018 from a 2-spot model to the ground-based data (middle row) and from a 4-spot model to the $K2$ C18 data (bottom row). The surface maps from left to right show the stellar hemispheres at primary minimum, first quadrature, secondary minimum and second quadrature, respectively.}
   \label{spots}
 \end{figure}

Figure~\ref{spots} shows the locations of the active regions on the giant star's surface in 2015 and 2018 at the phases of the minima and the quadratures. Note that only the longitudes and sizes of the spots are reliably determined (see above and Sect.\ref{5col}), and only the fixed latitude results are displayed.   

\subsection{Correlation between photometry and RV residuals}

\begin{figure}[tbp]
   \centering
      \includegraphics[width=9 cm]{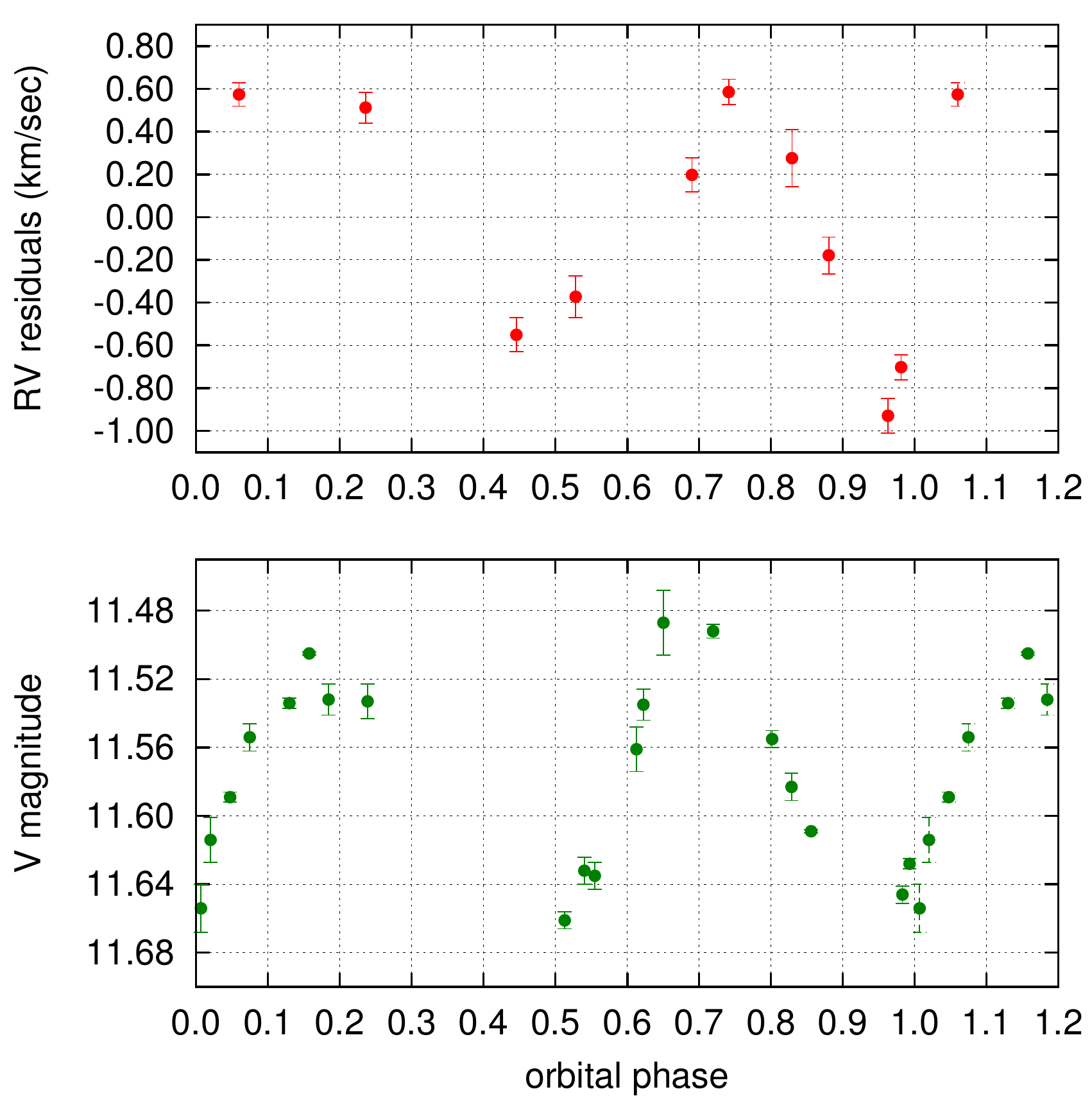}
   \caption{Comparison of the radial velocity residuals ({\it top}, red dots) with the nearly contemporaneous photometry ({\it bottom}, green dots).  The curves have similar shapes, probably reflecting a velocity jitter in the results caused by the spots.}
   \label{RV_phot_fit}
 \end{figure}

It is interesting to compare the radial velocity residuals with the light variations using the nearly contemporaneous RV and $BVR_CI_C$ photometry from 2018, since the effects of the spots can alter the radial velocities (referred to as `radial velocity jitter'). As shown by \"Ozdarcan et al. \cite{jitter} (see their Fig.~2.) the radial velocity residuals show the same rotational period as the star, and variations in the light curve are correlated with changes in the jitter curve, clearly demonstrating the effect of the spots. Looking at Fig.\ref{RV_phot_fit} we see a similar feature, i.e., that the radial velocity jitter of EPIC~211759736 (upper panel) follows a similar pattern and period as the rotational modulation in brightness caused by spots (lower panel). The full jitter amplitude is somewhat high, but strong activity, higher $v\mathrm{sin}\,i$ (19.6 km/sec in our case), and lower spectral resolution can all cause higher amplitude jitter. Many important details on how the radial velocity jitter appears in spotted stars can be found in Korhonen et al.~(\cite{heidi}).

\begin{table}[t]
\caption{Orbital elements of EPIC~211759736}             
\label{specorbit}      
\centering                          
\begin{tabular}{l l}       
\hline\hline \noalign{\smallskip} 
Parameter & value\\
 \noalign{\smallskip}\hline
 \noalign{\smallskip}
 P (days) & 36.522 $\pm$ 0.001 \\
 T$_0$ (BJD) & 2457168.181 $\pm$ 0.001 \\
 $\tau$ (HJD) & 2457149.82 $\pm$ 0.05 \\
a ($R_\odot$) & 63.8 $\pm$ 1.3 \\
e & 0.057 $\pm$ 0.003\\
$\omega$ (deg)& 269.2 $\pm$ 0.5\\
i (deg) & 81.85 $\pm$ 0.26\\ 
$\gamma$ (km/s) & 29.74 $\pm$ 0.08\\
$K_1$ (km/s) & 30.826\\
$K_2$ (km/s) & 57.257 \\ 
q & 0.54 $\pm$ 0.04\\
\noalign{\smallskip}\hline                  
\end{tabular}
\end{table}


\begin{table}[t]
\caption{Parameters of the components of EPIC\,211759736}             
\label{params}      
\centering                          
\begin{tabular}{l l l}       
\hline\hline \noalign{\smallskip} 
Parameter & Primary & Secondary\\
 \noalign{\smallskip}\hline
 \noalign{\smallskip}
Temperature, K & 4750 (adopted)& 5283$\pm$154\\
Radius, $R_\odot$ & 12.95$\pm$0.34 & 0.82$\pm$0.03\\
Mass, $M_\odot$ & 1.69$\pm$0.12 & 0.92$\pm$0.04\\
Luminosity, $L_\odot$ & 77.96$\pm$4.19 & 0.47$\pm$0.07\\
$M_V$, mag & 0.45$\pm$0.06 & 5.77$\pm$0.15\\
Gravity log $g/g_\odot$ & 2.44 & 4.58 \\
Metallicity & $-$0.27$\pm$0.13 & --- \\
\noalign{\smallskip}\hline                  
\end{tabular}
\end{table}

\section{Discussion}


\begin{table*}[t]
\caption{EPIC~211759736 and other binaries with giant components in the evolutionary tracks}             
\label{comp_params}      
\centering                          
\begin{tabular}{l l l l l l l l l}       
\hline\hline \noalign{\smallskip} 
Binary & [Fe/H] & log($T_{\rm eff,1}$) (K ) & log($L_1$) ($L_\odot$) & log($T_{\rm eff,2}$) (K)& log($L_2$) ($L_\odot$)& $M_{1,\odot}$ &  $R_{1,\odot}$ & ref.\\
\noalign{\smallskip}\hline\\ 
EPIC~211759736 & $-0.27\pm0.13$ & $3.677$ & $1.892\pm0.023$ & $3.723\pm0.013$ & $-0.331\pm0.066$ & $1.69\pm0.12$ & $13.0\pm0.3$ &1 \\
\noalign{\smallskip} 
KIC~4569590& $-0.34\pm0.09$ & $3.673\pm0.014$ & $1.942\pm0.062$ & $3.810\pm0.014$ & $0.157\pm0.070$ &$1.56\pm0.10$ & $14.1\pm0.2$ & 2\\
KIC~9540226& $-0.33\pm0.04$ & $3.671\pm0.006$ & $1.853\pm0.029$ & $3.806\pm0.006$ & $0.168\pm0.032$ & $1.33\pm0.05$& $12.8\pm0.1$ & 2\\
                      &                           &                              & $1.905\pm0.035^a$ &                               &                          & $1.45\pm0.05$ & $13.6\pm0.2$ & 2\\ 
XX~Tri           & $-0.27\pm0.03$ &$3.663$                & $1.452$                 &                               &                             & $1.26\pm0.15$ & $10.9\pm1.2$ & 3,6\\
                      &                           &$3.661\pm0.010$  & $1.532\pm0.006$  &                               &                             & & &4\\
$\zeta$~And  & $-0.30\pm0.05$ &$3.663\pm0.009$  & $1.98$                   &                               &                             & $2.6\pm0.4$ & $16.0\pm0.2$ & 5\\
                      &                           &$3.657\pm0.012$  & $1.984\pm0.002$  &                               &                             & & &4\\
BE~Psc         & 0.0                     & $3.653\pm0.007$  & $1.723\pm0.076$  & $3.799\pm0.007$ & $0.708\pm0.070$ &$1.56\pm0.03$ &$12.0\pm0.7$& 7\\                  
\noalign{\smallskip}\hline
\noalign{\smallskip}           
\end{tabular}
\tablefoot{$^a$: astroseismic value, 1: present work, 2: Gaulme et al. (\cite{gaulmeetal2}), 3: Ol\'ah et al. (\cite{overactive}), 4: {\it Gaia} DR2, 5: K\H ov\'ari et al. (\cite{zetaand}), 6: K\"unstler et al. (\cite{andreas}), 7: Strassmeier et al. (\cite{klausetal}).}
\end{table*}

It has been shown that some overactive K-giants in binaries, apart from their rotational modulation, exhibit cyclic, long-term light variations (`activity cycles') on the order of one magnitude in overall light variation due to strong and variable spot activity. The primaries of these binaries do not fit the theoretical evolutionary tracks, thereby resulting in irreconcilable age estimates (see, e.g., Ol\'ah et al. \cite{overactive}). The three stars studied in Ol\'ah et al. (\cite{overactive}), IL~Hya, XX~Tri and DM~UMa, are single-lined systems without eclipses, situated in the solar vicinity, and redder than the majority of giant stars in the solar neighborhood. 

Of these three stars XX~Tri has the same metallicity as EPIC~211759736, so it is sensible to compare them. Additionally, we choose two systems containing K-giant stars, KIC~9540226 and KIC~4569690, from Gaulme at al. (\cite{gaulmeetal1}, \cite{gaulmeetal2}) which have similar metallicities as our target and XX~Tri. These latter two systems have well-determined stellar parameters from the combined {\em Kepler} light curve and radial velocity analysis. One of them, KIC~9540226, in an eccentric orbit, does not show any observable spot activity. The giant star of this system has parameters both from asteroseismic and binary modeling, and was chosen because of its lack of magnetic activity for comparison with the more active stars. On the other hand, both components of KIC~4569690 seem to have spots, and the rotational modulation period of the primary is equal to the orbital period of the binary whose orbit is circular. The giant primary of this system does not show any asteroseismic signal and is the only one of the three systems in Gaulme et al.~\cite{gaulmeetal2} (their Table 1) with an active component which has similar metallicity as our target star. Finally, we took the well-studied $\zeta$~And (K\H ov\'ari et al. \cite{zetaand}), also with similar metallicity; this is the first active giant star having a direct interferometric image (Roettenbacher et al. \cite{rachael1}), which is a huge advantage when trying to understand the degree of spottedness. 

The properties of the four comparison stars and EPIC 211759736 are summarized in Table 6 where the parameters of the active eclipsing binary with giant primary BE ~Psc (of solar composition) are also given.

\begin{figure}[tbp]
   \centering
      \includegraphics[width=8.2cm]{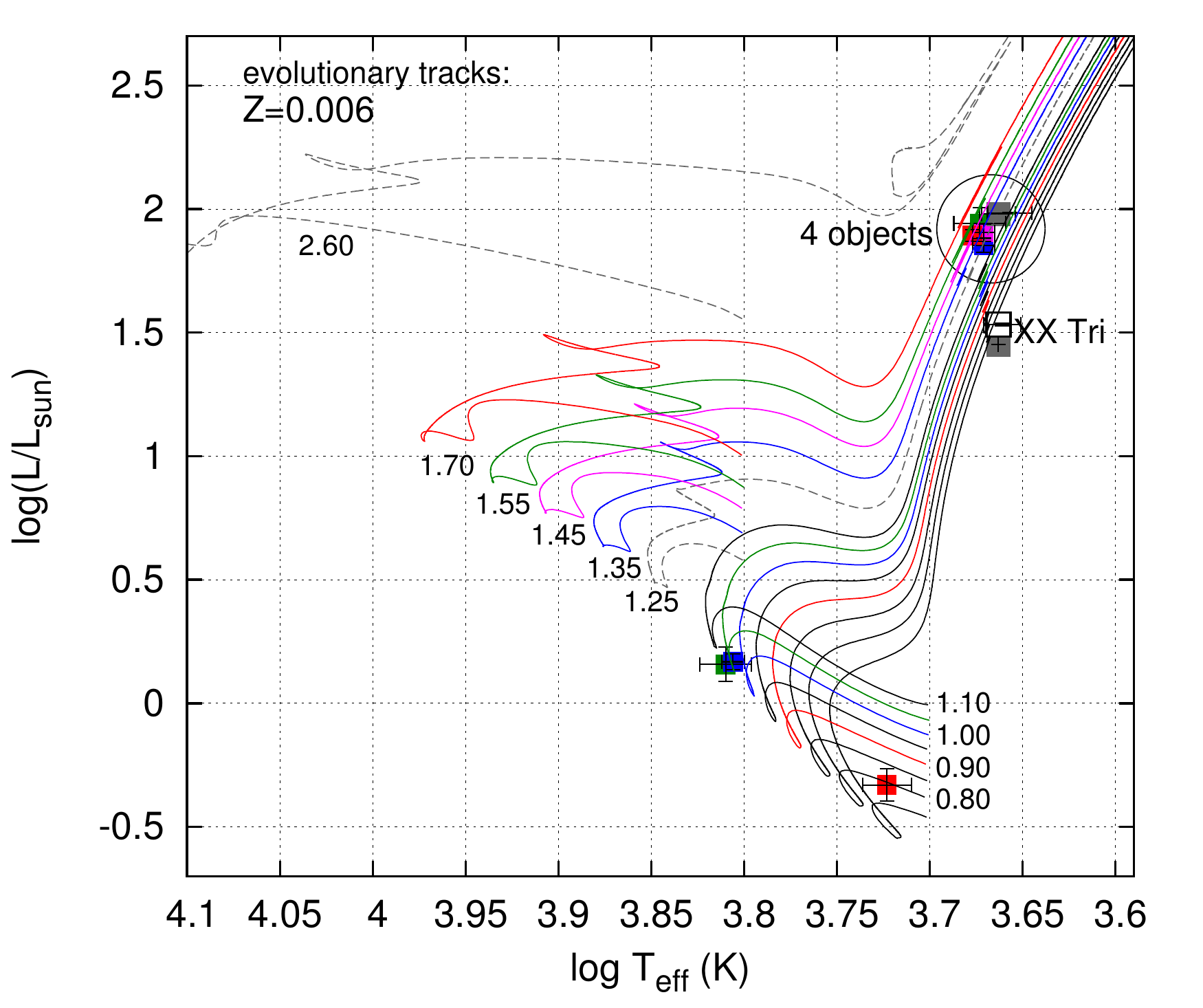}
      \includegraphics[width=8.2cm]{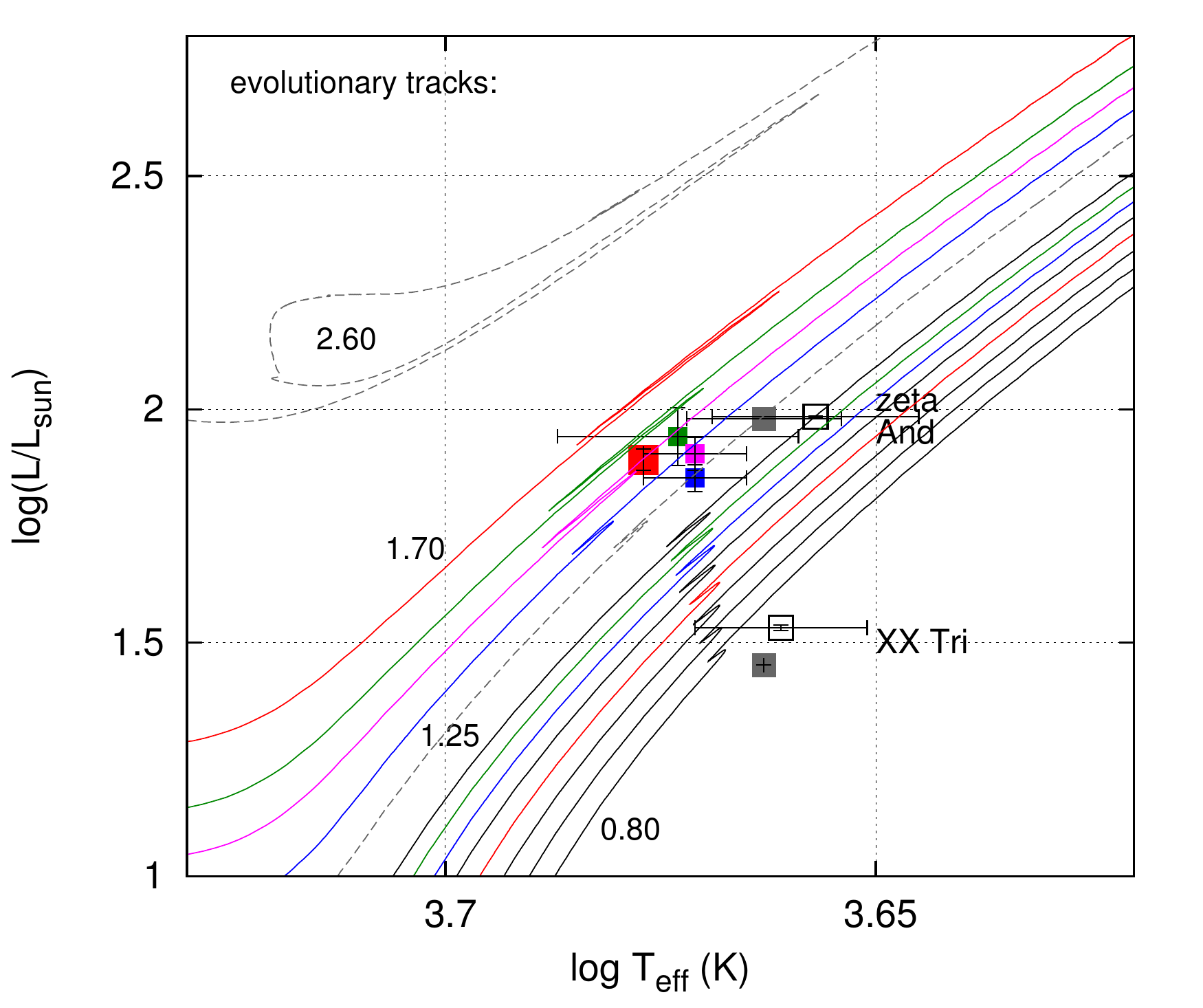}
      \includegraphics[width=8.2cm]{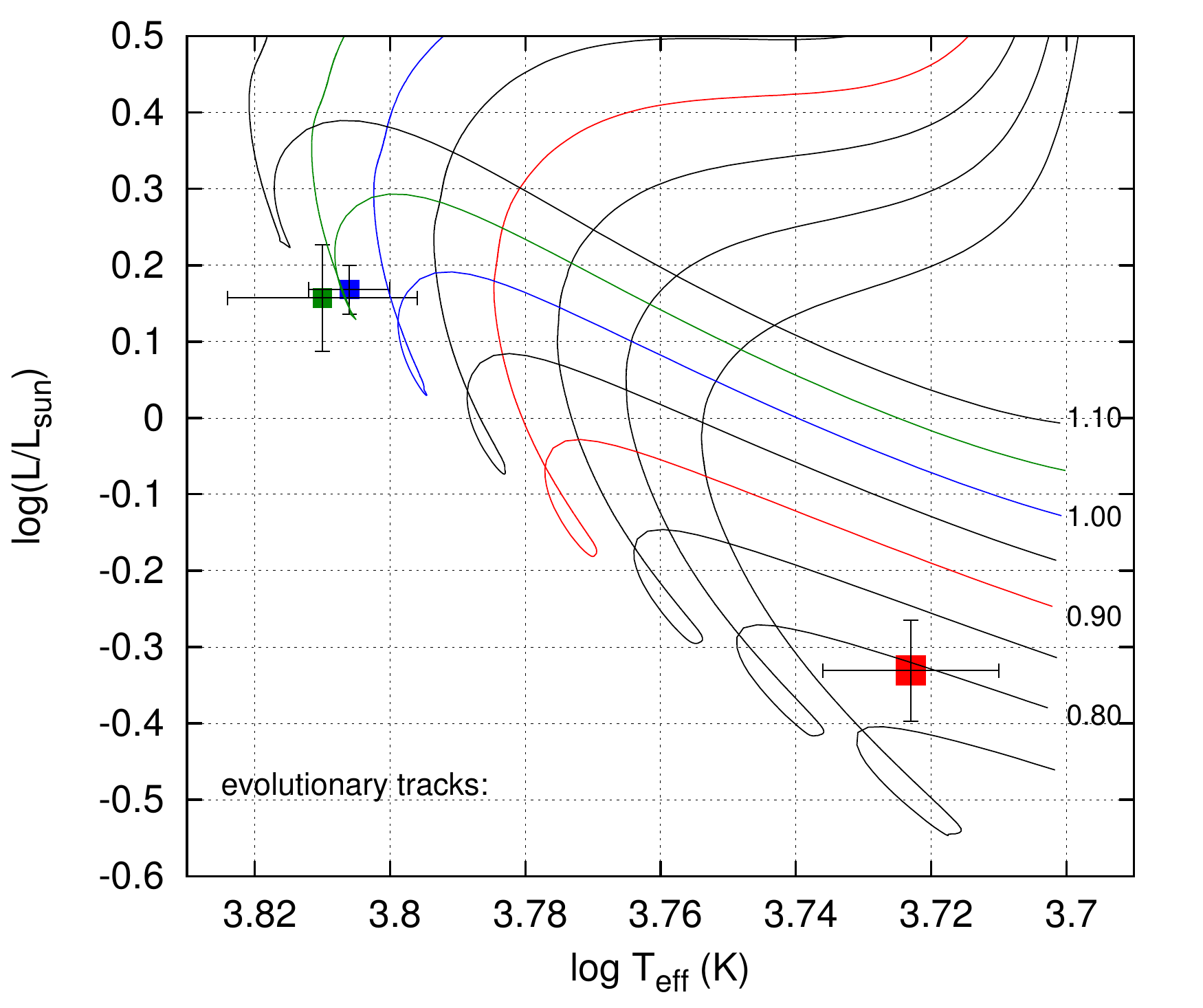}      
      \caption{Position of EPIC~211759736 (large red square) on the HRD (Bressan et al. \cite{bressan}), in comparison with KIC~4569690 (green) and KIC~9540226 (blue, dynamically derived values) and magenta (astroseismic values). Additionally, two well-known active giants are plotted: XX~Tri and $\zeta$~And with both the literature and {\it Gaia}2 results (filled and empty squares, respectively). Tracks for 2.6 and 1.25 solar masses are displayed in gray dashed lines for the traditionally determined $\zeta$~And and XX~Tri masses. The middle and bottom panels show the details around the giant and dwarf components of the systems. See the text and Table~\ref{comp_params} for details. }
   \label{tracks}
 \end{figure}

In Fig.\ref{tracks} we compare the positions of all five of these systems in the H-R diagram using the evolutionary tracks 
of Bressan et al. (\cite{bressan}). The plotted tracks belong to the metallicity of the studied stars, which are near to [Fe/H]$\sim-0.3$ (see Table~\ref{comp_params}). This translates to Z$\sim0.006$ with the standard abundances of Asplund et al. (\cite{asplund}). The star and the evolutionary track belonging to a given object are shown in the same color. In the case of XX~Tri and $\zeta$~And, the locations from both the literature (filled symbols) and from the {\it Gaia} results (open symbols, {\it Gaia} Collaboration, \cite{GAIA1}, \cite{GAIA2}) are plotted. For $\zeta$~And the two values are essentially the same, while for XX~Tri they are very close, emphasizing the reliability of the temperatures and luminosities derived from earlier observations (Table~\ref{comp_params}) of these two active giant stars. 

The masses of the secondary stars in the KIC systems match to within 1$\sigma$ their corresponding evolutionary tracks (Table~\ref{comp_params}). The secondary of EPIC~211759736 seems to be a bit too red for our determination that it is an unevolved MS star. However, this star is of solar type and may be found to show magnetic activity (as the primary does). That could result in a lower average temperature due to surface spots; for stellar temperature changes of active solar-type stars see Frasca \& Biazzo (\cite{fra_bia}).

Looking at Fig.~\ref{tracks} we see that all five giant primaries are situated below the tracks corresponding to their respective masses, though to differing degrees. In the middle panel of Fig.~\ref{tracks} the positions of EPIC~211759736 and the two KIC binaries are seen, enlarged. These latter two seem to be 1$\sigma$ below the corresponding evolutionary tracks for their respective masses, while the discrepancy is about 0.25~$M_\odot$ or 2$\sigma$ for EPIC~211759736. 

A very large discrepancy ($\sim$1.4~$M_\odot$) is is found for $\zeta$~And (K\H ov\'ari et al. \cite{zetaand}), which has large spots on its surface; its inclination is well constrained to be $70\pm2.8$ degrees  (Roettenbacher et al. \cite{rachael1}). However, the mass of $\zeta$~And was derived using such evolutionary tracks (for solar metallicities), which are now outdated. From the present evolutionary tracks on Fig.~\ref{tracks} the mass of $\zeta$~And is about 1.3~$M_\odot$, i.e., about half of its old value, and its age is around 3.5-4~Gyr; both of these sound reliable.
 
A similar discrepancy is seen for XX~Tri ($\gtrsim 0.5\,M_\odot$), as noted already in Ol\'ah et al. (\cite{overactive}), which has a long-term light variation with an amplitude of about one magnitude, and as well from time to time huge rotational modulations are present.  Large rotational modulations are observed only for stars with high rotational inclination---low inclinations or pole-on stars exhibit very small amplitude or no rotational modulation at all.  But on low-inclination objects it is still possible to observe large cyclic light variations, even though only about half of the stellar surface is visible to the observer (an example is V833~Tau, see Fig. 2. in Ol\'ah et al.~\cite{cycles}).

During the last decade stellar evolution calculations have become much more reliable (see the case of $\zeta$~And above), although the magnetic fields are still missing from the evolution models. The masses of the two KIC stars are not too discrepant from the theoretical value, and EPIC~211759736 is only 15\% higher in mass than implied by the evolutionary tracks. KIC~9540226 does not show light variations due to spots, and KIC~4569590 and EPIC~211759736 seem not to have strong activity which is inferred from their relatively small light-curve amplitudes originating from spots. The deviations from the direct modeling results for these three primary stars among the eclipsing binaries could be due to the still imperfect evolutionary models and/or different evolutions in binary systems. However, we do not have an independent mass determination for $\zeta$~And since it has no eclipses, but its light variation is also not very strong either in spot modulation or long-term timescale variations (cf. K\H ov\'ari et al. \cite{zetaand}). Therefore, very probably its recently determined mass is not too far from the correct one. 

\begin{figure}[tbp]
   \centering
      \includegraphics[width=8.5cm]{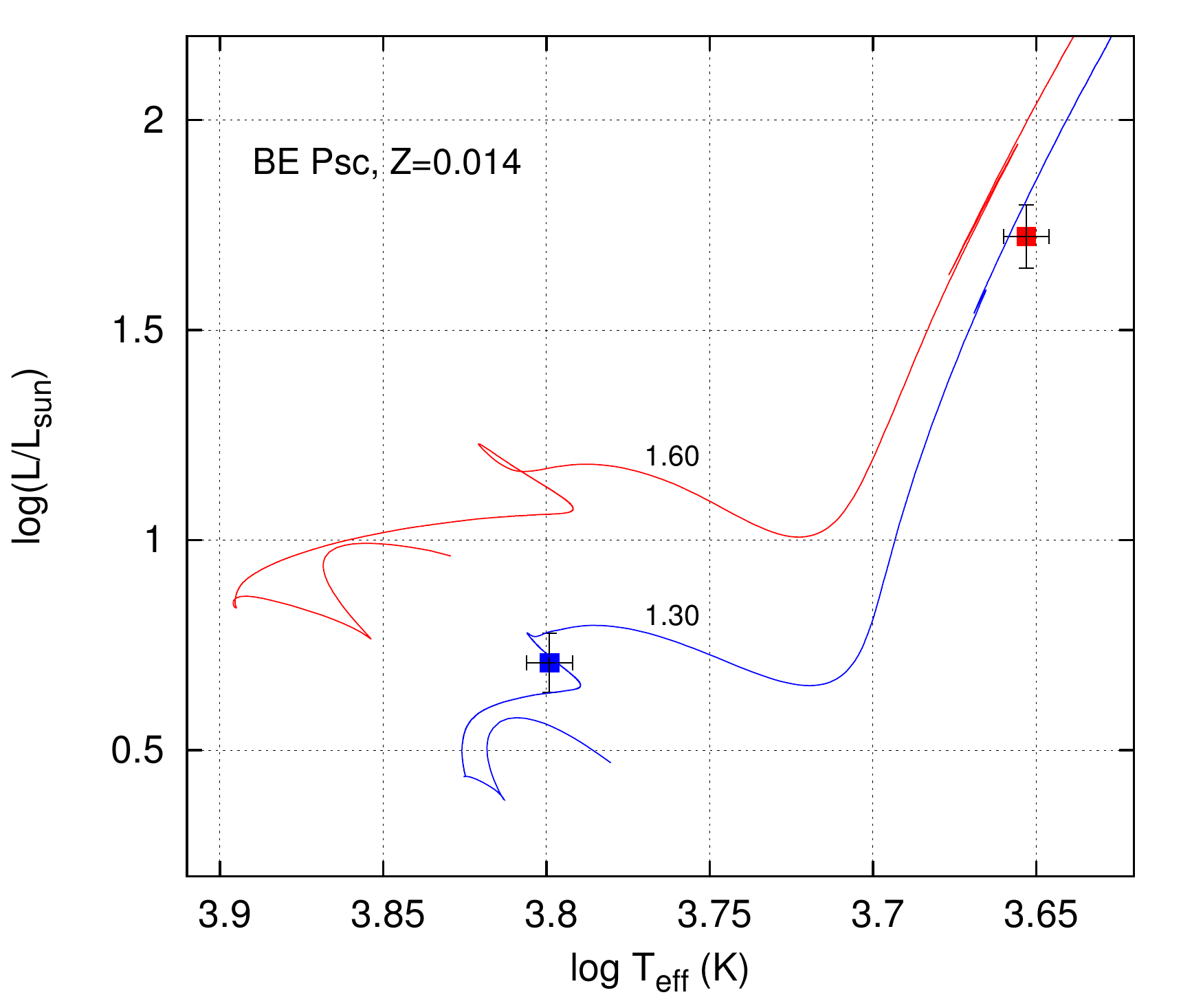}
            \caption{Position of BE~Psc on the HRD with metallicity close to solar. The positions of the primary  and secondary are marked with red and blue squares, respectively. The corresponding evolutionary tracks are plotted with the same color. While the secondary star matches its evolutionary track, slightly after having left the MS, the active primary deviates by about 30\% from its corresponding track (data are from Strassmeier et al. (\cite{klausetal}).}
   \label{bepsc}
 \end{figure}

The other well-studied active giant in an eclipsing binary, BE~Psc, has strong long-term and rotational variability (Strassmeier et al. \cite{klausetal}), and in the quoted paper the masses of the components are well determined from photometric and radial velocity data. The components are plotted on an HRD in Fig.~\ref{bepsc} for metallicity close to solar.  While the {\em secondary} stellar component is well matched by the evolutionary track corresponding to its known mass, the active {\em primary} deviates by about 30\% in the sense it has higher mass than its temperature and luminosity imply.

The case of XX~Tri is different than the others. Ol\'ah et al. (\cite{overactive}) did not find an acceptable mass for the star using {\it the same} evolutionary tracks as in the present paper; the derived temperature and luminosity point to an inconsistent mass and age on the HRD. This star is `overactive' with huge rotational modulations and long-term variability, and in this case we believe that the strong magnetic field was able to alter even its stellar structure. Present stellar evolution models that do not take into account the magnetic field are unable to predict a reasonable mass for XX~Tri. We note here that in the case of another `overactive' star, IL~Hya, weak evidence is found for changes in its stellar radius during the long-term cycle (Ol\'ah et al. \cite{overactive}). Even the question of radius changes in the well-observed nearest star, the Sun, is still open (cf. Kosovichev \& Rozelot \cite{kos_roz}).

The internal structure of giant stars with strong magnetic fields between the core and the widened, diluted atmosphere is not well studied. Supposing a flux tube dynamo scenario, the flux ropes created by the magnetic dynamo in the shear layer between the core and the convection zone rise up from the tachocline to the surface, causing the observable activity features, but can also remain trapped below the stellar surface (Holzwarth \& Sch\"uessler, \cite{volkmar}), and the consequences are unknown. The oscillations of some red giant stars are suppressed to undetectable level due to internal magnetic fields (Gaulme et al. \cite{gaulmeetal2}), but the mechanism for this is not completely clear.

\section{Summary and Conclusions}

In this paper we studied in detail EPIC~211759736, which is now the second best-studied/characterized active giant in an eclipsing binary after BE~Psc. We determined the physical parameters of both stellar components and provide a description of the rotational and long-term activity of the primary component. The temperatures and luminosities of both components were examined in the context of the HRD. We find that both the primary and the secondary components deviate from the evolutionary tracks corresponding to their masses in the sense that the stars appear in the HRD at lower masses than their true masses.

We compared EPIC~211759736 with the well-studied active giants BE~Psc, $\zeta$~And and XX~Tri.  Among these, only BE~Psc has eclipses.  Two KIC stars were also in the comparison sample. Except for BE~Psc,  all the stars have very similar metallicities. We suggest three possible reasons for the inferred mass deviations, if any, from the evolutionary tracks: inexact evolution calculations for evolved stars,
the difference in stellar evolution when a star is in a binary system, and the effect of the strong magnetic field on the stellar structure. 
Our results suggest that in the case of no/low activity, the evolutionary tracks agree fairly well with the derived masses of the stars: these are the two KIC stars, one without any observable activity and the other with low amplitude variability. Next, the target of our present study, EPIC~211759736, shows a higher amplitude rotational modulation and also long-term variability.  And, in this case, the primary is about 15\% more massive than is implied by the evolutionary tracks. BE~Psc has even stronger activity, and its primary component is about 30\% more massive than the evolutionary tracks suggest. Finally, XX~Tri has by far the strongest activity, but it lacks eclipses, so only the mass function is determinable. In the case of XX~Tri it is not possible to get a reliable mass from the HRD. 

The mass discrepancies seem to be larger as the strength of the stellar activity grows in our sample of active giant stars. Possibly the lack of magnetic fields in the evolution calculation and the effects of evolving in a binary play a role in their position on the HRD for all of our stars, but these would seem to be minor effects. The strength of the magnetic activity, however, may significantly alter even the structure of the stars. The KIC stars and $\zeta$~And possibly have weaker internal magnetic fields, while EPIC~211759736, BE~Psc and XX~Tri have progressively stronger magnetic fields revealed by their photometric behaviour. This is manifested clearly in the deviations of their masses from the theoretical predictions.

Observations of eclipsing binaries with active giant components is of crucial importance in describing stellar evolution after the MS in the presence of magnetic fields. EPIC~211759736 is now one of the two well-studied active giants in eclipsing binaries which should lead to better constraints on the structure and evolutionary trends for this type of magnetically active star.

Eclipsing spotted giant stars, such as EPIC~211759736, could benefit quite a bit from long-term, multicolor observations.  These would allow us to study the average temperature changes of the active regions, which may then reflect the possible spot/plage ratio changes. More effort is needed to further characterise systems such as EPIC~211759736 and BE~Psc, as well as most of the active  KIC stars from Gaume et al. (\cite{gaulmeetal2}), all of which have primary components between 1.0-1.6~$M_{\odot}$ and secondaries between 0.8-1.3~$M_{\odot}$. The different evolutionary stages of similar mass, solar-like stars in close binaries, with different activity levels of possibly magnetic origin in one or both components, should shed light on the effect of the magnetic field during the evolution on the MS and on the giant branch of the HRD.

\begin{acknowledgements}
Thanks are due to an anonymous referee for careful reading the the manuscript and for good suggestions.
We are grateful for H. Korhonen for advice concerning radial velocity jitter in active stars.  DL and TJ gratefully acknowledge Allan R. Schmitt for making his lightcurve examining software {\tt LcTools} freely available.
This work has been supported by the Hungarian Science Research Program OTKA-K-113117.
This publication makes use of data products from the Two Micron All Sky Survey, which is a joint project of the University of Massachusetts and the Infrared Processing and Analysis Center/California Institute of Technology, funded by the National Aeronautics and Space Administration and the National Science Foundation. This publication makes use of data products from the Wide-field Infrared Survey Explorer, which is a joint project of the University of California, Los Angeles, and the Jet Propulsion Laboratory/California Institute of Technology, funded by the National Aeronautics and Space Administration. The DASCH project at Harvard is grateful for partial support from NSF grants AST-0407380, AST-0909073, and AST-1313370.

\end{acknowledgements}

\end{document}